\documentclass[a4paper,11pt]{article}
\pdfoutput=1
\usepackage{jheppub}
\usepackage[english]{babel}
\usepackage{cite}

\usepackage{amsfonts}
\allowdisplaybreaks
\usepackage[mathscr]{euscript}
\usepackage{dsfont}
\usepackage{bbm}
\DeclareMathAlphabet{\mathdsl}{U}{bbm}{m}{sl}

\usepackage{tikz}
\usetikzlibrary{arrows}
\usetikzlibrary{calc}
\tikzset{>=latex}

\DeclareMathOperator{\tr}{tr}

\newcommand{\s}{$\sigma$}

\newcommand{\dd}{\mathrm{d}}
\newcommand{\fg}{\mathfrak{g}}
\renewcommand{\L}{\mathcal{L}}
\newcommand{\J}{\mathcal{J}}
\renewcommand{\th}{\widehat{t}}
\newcommand{\Fh}{\widehat{F}}
\newcommand{\etah}{\widehat{\eta}}
\newcommand{\Eh}{\widehat{\mathcal{E}}}
\newcommand{\sEh}{\widehat{E}}
\newcommand{\Th}{\widehat{T}}
\newcommand{\betah}{\widehat{\beta}}
\newcommand{\Ph}{\widehat{P}}
\newcommand{\Pbh}{\widehat{\overline{P}}{}}

\newcommand{\cg}{c_{\mathfrak{g}}}
\newcommand{\EE}{\mathcal{E}}
\newcommand{\dlie}{\mathfrak{d}}
\newcommand{\hlie}{\mathfrak{h}}
\newcommand{\DD}{\mathds{D}}
\newcommand{\HH}{\mathds{H}}
\def\res{\mathop{\text{res}\,}}
\newcommand{\Mh}{\widehat{M}}
\newcommand{\vh}{\hat{v}}
\newcommand{\SSh}{\widehat{\mathcal{S}}}

\newcommand{\Ah}{\widehat{A}}
\newcommand{\Bh}{\widehat{B}}
\newcommand{\Ch}{\widehat{C}}
\newcommand{\Dh}{\widehat{D}}

\newcommand{\ah}{\widehat{a}}
\newcommand{\bh}{\widehat{b}}
\newcommand{\ch}{\widehat{c}}
\renewcommand{\dh}{\widehat{d}}
\newcommand{\eh}{\widehat{e}}
\newcommand{\fh}{\widehat{f}}
\newcommand{\ahb}{\overline{\widehat{a}}}
\newcommand{\bhb}{\overline{\widehat{b}}}
\newcommand{\chb}{\overline{\widehat{c}}}

\newcommand{\ab}{\bar a}
\newcommand{\bb}{\bar b}
\newcommand{\cb}{\bar c}

\newcommand{\MS}{\mathrm{MS}}
\newcommand{\CM}{\mathrm{CM}}

\title{\boldmath The Magic Renormalisability of Affine Gaudin Models}

\author[a]{Falk Hassler,}
\author[b]{Sylvain Lacroix,}
\author[c]{Beno\^{\i}t Vicedo}

\emailAdd{falk.hassler@uwr.edu.pl}
\emailAdd{sylvain.lacroix@eth-its.ethz.ch}
\emailAdd{benoit.vicedo@gmail.com}

\affiliation[a]{Faculty of Physics and Astronomy, University of Wrocław, Maksa Borna 9, 50-204 Wrocław, Poland}
\affiliation[b]{Institute for Theoretical Studies, ETH Z\"urich, Clausiusstrasse 47, 8092 Z\"urich, Switzerland}
\affiliation[c]{Department of Mathematics, University of York, York YO10 5DD, United Kingdom}

\abstract{We study the renormalisation of a large class of integrable $\sigma$-models obtained in the framework of affine Gaudin models. They are characterised by a simple Lie algebra $\mathfrak{g}$ and a rational twist function $\varphi(z)$ with simple zeros, a double pole at infinity but otherwise no further restrictions on the pole structure. The crucial tool used in our analysis is the interpretation of these integrable theories as $\mathcal{E}$-models, which are $\sigma$-models studied in the context of Poisson-Lie T-duality and which are known to be at least one- and two-loop renormalisable. The moduli space of $\mathcal{E}$-models still contains many non-integrable theories. We identify the submanifold formed by affine Gaudin models and relate its tangent space to curious matrices and semi-magic squares. In particular, these results provide a criteria for the stability of these integrable models under the RG-flow. At one loop, we show that this criteria is satisfied and derive a very simple expression for the RG-flow of the twist function, proving a conjecture made earlier in the literature.}

\begin{document}

\maketitle

\section{Introduction}
A major challenge of contemporary physics is to study quantum field theories at strong coupling. An arguably important example is quantum chromodynamics (QCD) at low energies. Here, the usually very successful perturbative treatment breaks down and new strategies are required. One of them is to exploit symmetries to obtain predictions of strong coupling phenomena. This works better the more restrictive the considered symmetries are. According to this criterion, the clear winners are integrable systems, whose dynamics are completely fixed by the existence of infinitely many conserved charges. Although QCD is not integrable, it has an integrable relative, namely the maximally supersymmetric SU($N$) Yang-Mills theory, which shares some of its qualitative features. This theory has been actively analysed with integrability techniques over the past two decades \cite{Beisert:2010jr}. Most notably, it is equivalent to a closed string propagating in five-dimensional Anti-de Sitter (AdS) space by the AdS/CFT correspondence. In agreement with this correspondence, the integrability is also manifest in the two-dimensional non-linear \s-model which describes the string propagating in the AdS space. Therefore, integrable \s-models currently attract considerable interest. They promise an unprecedented level of computational control but are at the same time very hard to find and even harder to quantise.

In this article, we consider a large family of integrable field theories called affine Gaudin models~\cite{Levin:2001nm,Feigin:2007mr,Vicedo:2017cge}, which gained particular attention in the recent years because they provide a unifying framework encompassing most of the known integrable \s-models as well as many new examples~\cite{Delduc:2018hty,Delduc:2019bcl,Lacroix:2019xeh}. Unfortunately, the standard approach to quantise integrable field theories, the inverse scattering method, is not directly applicable to affine Gaudin models due to the so-called problem of non-ultralocality~\cite{Takhtajan:1979iv,Faddeev:1979gh}. However, there are alternatives to probe their quantum regime. For example, one might ask how counter terms, that originate from loop corrections, affect integrability. Do they break classical integrability or not? We take this question as the motivation for this article. To avoid getting lost in technicalities, we further restrict the discussion to models without supersymmetry or additional gauge symmetries in the realm of bosonic string theory. They are captured by the bosonic string \s-model, namely\footnote{Later on, we parameterise the worldsheet by two coordinates $\tau$ and $\sigma$, where the former is timelike. The worldsheet metric is chosen such that $\star 1 = \dd\tau\wedge\dd\sigma$, $\star \dd\tau = \dd \sigma$, $\star \dd\sigma=\dd\tau$ and $\star( \dd\tau\wedge\dd\sigma) = -1$.}
\begin{equation}\label{eqn:smodel}
  S = \frac1{4\pi\alpha'} \int_\Sigma \left( g_{ij} \, \dd x^i \wedge \star \dd x^j + B_{ij} \, \dd x^i  \wedge \dd x^j \right) \,.
\end{equation}
It governs $D$ scalars $x^i$, $i=1,\dots,D$, which capture the embedding of the closed string into the $D$-dimensional (pseudo)-Riemannian target space with the metric $d s^2 = g_{ij} \,\dd x^i \dd x^j$ and the two-form gauge potential $B = \frac12 B_{ij}\, \dd x^i \wedge \dd x^j$. Only for very few distinguished choices of the metric and the $B$-field this model becomes an integrable field theory. Finding them is in general a difficult task as there is no completely systematic procedure known yet (if it even exists) to determine whether a given choice of target space geometry leads to an integrable dynamics or not. However, the formalism of affine Gaudin models allows the systematic construction of a very large class of such integrable \s-models, which encompasses most known examples. Therefore, we choose to focus on this class here.

Quantum corrections will renormalise both $g_{ij}$ and $B_{ij}$. It is by no means clear that they should stay in the very small subset of metrics and $B$-field that result in an integrable \s-model. Therefore, it is natural to ask: 
\begin{center}
  Is integrability preserved (or stable) under renormalisation group (RG)-flows?
\end{center}
The main challenge in finding an answer is that we lack a simple criteria to decide if a given combination of metric and $B$-field describes an integrable \s-model or not. The only currently feasible approach is to compare the result of the flow with known examples of integrable models. This has been done successfully over the last years \cite{Itsios:2014lca,Sfetsos:2015nya,Georgiou:2017jfi,Demulder:2017zhz,Hoare:2019ark,Hoare:2019mcc,Delduc:2020vxy,Levine:2021fof,Levine:2022hpv,Levine:2023wvt} providing clues that integrability is stable under RG-flows. This finding is particularly remarkable, because at the level of the \s-model \eqref{eqn:smodel}, it is not obvious at all and seems rather magic. Of course there is no magic but rather a hidden symmetry principle that protects the model from integrability breaking counter terms. Our objective here is to reveal it for affine Gaudin models.

To this end, we first discuss in section \ref{sec:PLsym} a larger class of (not necessarily integrable) \s-models, called $\EE$-models~\cite{Klimcik:1995dy,Klimcik:1996nq,Klimcik:2015gba}, and review the properties of their symmetries, dualities and renormalisation. In section \ref{sec:integrability}, we introduce affine Gaudin models as a specific subclass of $\EE$-models which are automatically integrable, as their construction ensures the existence of a Lax connection and a Maillet $r$/$s$ algebra \cite{Maillet:1985ek, Maillet:1985ec}. Their formulation in terms of $\EE$-models will be extremely useful when we finally transition from the classical to the quantum regime in section~\ref{sec:betafuncs}. More specifically, we will investigate the form of quantum corrections for affine Gaudin models. At this point we will identify the structure which preserves integrability under the RG-flow as related to semi-magic matrices. Moreover, we will use these results to establish the 1-loop renormalisability of these models and to prove a conjecture formulated in \cite{Delduc:2020vxy}, which compactly expresses their 1-loop RG-flow in terms of their so-called twist function. This generalises the results of \cite{Hassler:2020xyj} to the case where the twist function possesses poles of arbitrary order in the complex plane.

\section{\texorpdfstring{$\EE$}{E}-models and Poisson-Lie Symmetry\label{sec:PLsym}}
\subsection{\texorpdfstring{$\EE$}{E}-models in terms of current algebras}
In this section, we review the general formalism of $\EE$-models~\cite{Klimcik:1995dy,Klimcik:1996nq,Klimcik:2015gba} and their relations to Poisson-Lie symmetry \cite{Klimcik:1995ux}. Our starting point is a Lie algebra $\dlie$, with basis $\lbrace \th_{\Ah} \rbrace$ (where the hatted notations are introduced for future convenience) and structure constants $\Fh_{\Ah\Bh}{}^{\Ch}$ defined by the bracket
\begin{equation}\label{eqn:liealgebra}
  [ \th_{\Ah}, \th_{\Bh} ] = \Fh_{\Ah\Bh}{}^{\Ch}\, \th_{\Ch}\,.
\end{equation}
We assume that $\dlie$ is equipped with an ad-invariant non-degenerate symmetric pairing
\begin{equation}
  \langle \th_{\Ah}, \th_{\Bh} \rangle = \etah_{\Ah\Bh}\,,
\end{equation}
which requires
\begin{equation}
 \Fh_{\Ah\Bh}{}^{\Dh} \, \etah_{\Dh\Ch} + \Fh_{\Ah\Ch}{}^{\Dh}\, \etah_{\Bh\Dh} = 0\,.
\end{equation}
Finally, we suppose that we are given a symmetric involution $\Eh: \dlie \rightarrow \dlie$, $\Eh^2 = \mathrm{id}$, such that its two eigenspaces $\text{Ker}(\Eh \pm \mathrm{id})$ have the same dimension. The operator $\Eh$ acts on the basis $\lbrace \th_{\Ah} \rbrace$ as
\begin{equation}\label{eqn:Eop}
  \Eh ( \th_{\Ah} ) = \th_{\Bh}\, \Eh^{\Bh}{}_{\Ah}.
\end{equation}
We will use the pairing $\etah_{\Ah\Bh}$ and its inverse $\etah^{\,\Ah\Bh}$ to lower and raise indices. In particular, the ad-invariance of this pairing and the symmetry of $\Eh$ imply
\begin{equation}
\Fh_{\Ah\Bh\Ch} = - \Fh_{\Ah\Ch\Bh} \quad \text{ and } \quad \Eh_{\Ah\Bh} = \Eh_{\Bh\Ah}\,.
\end{equation}
Together with $\Fh_{\Ah\Bh\Ch} = - \Fh_{\Bh\Ah\Ch}$, which follows from its origin as structure coefficients of the Lie algebra \eqref{eqn:liealgebra}, this implies that $\Fh_{\Ah\Bh\Ch}$ is totally anti-symmetric.

If $\etah_{\Ah\Bh}$ has split signature, i.e. the same number of positive and negative eigenvalues, this data describes an $\EE$-model. The latter has been introduced to study Poisson-Lie symmetric \s-models and the closely related Poisson-Lie T-duality \cite{Klimcik:1995ux}. The name Poisson-Lie originates from the fact that the first examples of Lie algebras $\dlie$ studied in this context have been Drinfeld doubles describing Lie bialgebras associated to Poisson-Lie groups.

The $\EE$-model is defined as a Hamiltonian field theory with periodic spatial coordinate $\sigma\in[0,2\pi)$ whose fundamental field is a $\dlie$-valued current $\J_\sigma(\sigma)=\th_{\Ah}\,\J_\sigma^{\Ah}(\sigma)$ satisfying the Poisson bracket \cite{Klimcik:2015gba}
\begin{equation}\label{eqn:PBJh}
  \bigl\{\J_\sigma^{\Ah}(\sigma_1), \J_\sigma^{\Bh}(\sigma_2)\bigr\} = 
    \Fh^{\Ah\Bh}{}_{\Ch}\, \J_\sigma^{\Ch}(\sigma_1)\, \delta(\sigma_1-\sigma_2) - \etah^{\,\Ah\Bh}\, \delta'(\sigma_1-\sigma_2)\,.
\end{equation}
Evolution with respect to the time coordinate $\tau$ of the model of any observable is fixed by $\partial_\tau = \lbrace H,\cdot \rbrace$, where the Hamiltonian is chosen as
\begin{equation}\label{eqn:EmodelH}
  H = \frac12 \int_0^{2\pi} \dd\sigma \, \Eh_{\Ah\Bh}\, \J_\sigma^{\Ah} \J_\sigma^{\Bh} = \frac12 \int_0^{2\pi} \dd\sigma \,  \langle \J_\sigma, \Eh(\J_\sigma) \rangle\,.
\end{equation}
All field equations induced by this Hamiltonian can be combined into the compact form
\begin{equation}
  \dd \J^{\Ah} + \J^{\Bh} \wedge \J^{\Ch}\, \Fh_{\Bh\Ch}{}^{\Ah} = 0
\end{equation}
based on the $\dlie$-valued 1-form $\J = \th_{\Ah} \J^{\Ah} = \J_\tau \dd \tau + \J_\sigma \dd \sigma$, if one identifies
\begin{equation}
  \J_\tau = \Eh( \J_\sigma) = \th_{\Bh}\,\Eh^{\Bh}{}_{\Ah}\J_\sigma^{\Ah}\,. 
\end{equation}

\subsection{\texorpdfstring{\s}{sigma}-model realisations and isotropic subalgebras}
\label{sec:sigma}
There are several ways of connecting the $\EE$-model defined above with the standard \s-model \eqref{eqn:smodel}. All of them require choosing a maximally isotropic subalgebra $\hlie \subset \dlie$. Maximality implies that the dimension of $\hlie$ is half the dimension of $\dlie$, while the isotropy requires that the pairing between any pair of $\hlie$'s generators vanishes. For every choice of such $\hlie$, one can construct a realisation of the $\EE$-model as a \s-model whose target space is the quotient $M=\HH\backslash\DD$, where $\HH$ and $\DD$ are the Lie groups corresponding to $\hlie$ and $\dlie$. In order to make this relation explicit, we have to explain how the fundamental fields of the $\EE$-model, namely the current components $\bigl( \J_\sigma^{\Ah}(\sigma) \bigr)_{\Ah=1,\dots,\,\text{dim} \,\dlie}$, are related to the degrees of freedom of the \s-model. As above, we will work in the Hamiltonian formulation: the \s-model is then described by coordinate fields $\bigl( x^i(\sigma) \bigr)_{i=1,\dots,\,\text{dim} \,M}$ on the target space $M$ and their conjugate momenta $\bigl( p_i(\sigma)\bigr)_{i=1,\dots,\,\text{dim}\, M}$, satisfying the canonical Poisson brackets
\begin{equation}
\{ p_i(\sigma_1), x^j(\sigma_2) \} = \delta_i{}^j\,\delta(\sigma_1-\sigma_2) \qquad \{ x^i(\sigma_1), x^j(\sigma_2) \} = \{ p_i(\sigma_1), p_j(\sigma_2) \} = 0\,.
\end{equation}
Of particular interest for us will be the collection of fields
\begin{equation}
\big( \J^I_\sigma(\sigma) \big)_{I = 1, \ldots, 2M} = \left( \partial_\sigma x^i(\sigma), p_i(\sigma) \right)_{i = 1, \ldots, M}\,,
\end{equation}
which is naturally valued in the generalised tangent bundle $T M \oplus T^* M$. In the above equation and in what follows, we use upper-case indices $I,J \in \{ 1,\dots, 2\text{ dim}\,M \}$ to label components in this bundle, in contrast with the lower-case indices $i,j\in \{ 1,\dots, \text{ dim}\,M \}$ labelling coordinates on $M$. The fields $\J^I_\sigma$ are governed by the bracket
\begin{equation}
\{\J_\sigma^I(\sigma_1), \J_\sigma^J(\sigma_2)\} = - \etah^{\,IJ}\, \delta'(\sigma_1-\sigma_2)\,, \qquad \text{where} \qquad \etah^{\,IJ} = \begin{pmatrix}
        0 & \delta^i{}_j \\
        \delta_i{}^j & 0
      \end{pmatrix}
\end{equation}
is the standard pairing on the generalised tangent bundle $T M \oplus T^* M$. As commonly done, we use $\etah^{\,IJ}$ and its inverse $\etah_{\,IJ}$ to raise and lower indices $I,J,\dots$\,. The notations $\J^I_\sigma$ and $\etah^{\,IJ}$, similar to the current components $\J^{\Ah}_\sigma$ and bilinear pairing $\etah^{\,\Ah\Bh}$ on $\dlie$, are justified by the pivotal formula
\begin{equation}\label{eqn:Jsigma}
  \J^{\Ah}_\sigma := \sEh^{\Ah}{}_I \,\J^I_\sigma\,,
\end{equation}
which relates the $\EE$-model current to the \s-model fields. The key ingredient in this relation is the so-called generalised frame field $\sEh^{\Ah}{}_I$, which satisfies the following properties:
\begin{enumerate}
  \item It depends only on the coordinates $x^i$ of the quotient $M=\HH\backslash \DD$.
  \item It transforms the standard metric $\etah^{\,IJ}$ on $T M \oplus T^* M$ into the pairing $\etah^{\,\Ah\Bh}$ on $\dlie$ , \textit{i.e.}
    \begin{equation}
     \etah^{\,\Ah\Bh} = \etah^{\,IJ}\, \sEh^{\Ah}{}_I\, \sEh^{\Bh}{}_J .\vspace{-4pt}
    \end{equation}
  \item It gives rise to the structure coefficients through the relation
    \begin{equation}
      \Fh_{\Ah\Bh\Ch} = 3 \sEh_{[\Ah}{}^I \partial_I \sEh_{\Bh}{}^J \sEh_{\Ch]J}\,,
    \end{equation}
    where $\partial_I = \big( \partial_i \; 0 \big)$ is the standard solution of the section condition in double field theory \cite{Siegel:1993th,Hull:2009mi,Aldazabal:2013sca}.
\end{enumerate}
The existence of a generalised frame field with such properties is a non-trivial fact \cite{Hassler:2017yza,Demulder:2018lmj,Sakatani:2019jgu,Hassler:2019wvn,Borsato:2021vfy}, which relies strongly on the assumption that $\hlie$ is a maximally isotropic subalgebra of $\dlie$. More precisely, the explicit construction of $\sEh^{\Ah}{}_I$ extensively uses the so-called isotropic basis of $\dlie$ associated with the choice of $\hlie$. For completeness, we review this in Appendix \ref{app:Frame}. The properties 1 to 3 listed above ensure that, through the relation \eqref{eqn:Jsigma}, the canonical bracket for the fields of the \s-model translates to the current algebra \eqref{eqn:PBJh} for $\J^{\Ah}_\sigma$. This guarantees that the formula \eqref{eqn:Jsigma} is compatible with the Poisson structure underlying the $\EE$-model.

With this relation at hand, one is able to reformulate the $\EE$-model entirely in terms of the canonical fields $(x^i,p_i)$. In particular, the Hamiltonian \eqref{eqn:EmodelH} becomes
\begin{equation}\label{eqn:Hsigma}
  H = \frac{1}{2} \int_0^{2\pi} \dd\sigma\, \Eh_{IJ}\, \J_\sigma^{I} \J_\sigma^{J}
    \qquad\text{with}\qquad
  \Eh_{IJ} = \Eh_{\Ah\Bh} \, \sEh^{\Ah}{}_I \, \sEh^{\Bh}{}_J\,. \end{equation}
The object $\Eh_{IJ}$ appearing in this formula is called the generalised metric and takes the form \cite{Tseytlin:1990nb}
\begin{equation}\label{eqn:EtogandB}
  \Eh_{IJ} =
  \begin{pmatrix}
    g_{ij} - B_{ik} g^{kl} B_{lj} \quad& B_{ik} g^{kj} \\
    -g^{ik} B_{kj} & g^{ij}
  \end{pmatrix}\,,
\end{equation}
where the tensors $g_{ij}=g_{ji}$, $g^{ij}=g^{ji}$ and $B_{ij}=-B_{ji}$ on $M$ are identified with the metric, inverse metric and $B$-field of the \s-model \eqref{eqn:smodel}, respectively. Indeed, a standard canonical analysis shows that \eqref{eqn:Hsigma} is exactly the Hamiltonian corresponding to the action $2 \pi \alpha' S$. In particular, we can conveniently read off the metric and $B$-field from $\Eh_{IJ}$. All challenges in the identification of the $\EE$-model with a \s-model are encapsulated in the generalised frame field, but fortunately its construction is completely worked out (see above and Appendix \ref{app:Frame}). If there are different possible choices for the maximally isotropic subgroup $\HH$, we can construct different generalised frame fields that are not related to each other by diffeomorphisms or $B$-field gauge transformations. Thus, they correspond to different \s-models which still arise from the same $\EE$-model. This phenomena is known as Poisson-Lie T-duality.

To make the discussion complete, let us finally spell out the relation
\begin{equation}
    p_i = g_{ij}\,\partial_\tau x^j + B_{ij}\,\partial_\sigma x^j
\end{equation}
between the conjugate momenta $p_i$ used in the Hamiltonian formulation of the \s-model and the time derivatives $\partial_\tau x^i$ appearing at the Lagrangian level. Recall that the dynamics of the $\EE$-model is conveniently expressed in terms of the 1-form $\J^{\Ah} = \J^{\Ah}_\sigma\,\dd \sigma + \J^{\Ah}_\tau\,\dd \tau$. The equation \eqref{eqn:Jsigma} above explained the relation between its spatial component $\J^{\Ah}_\sigma$ and the canonical fields of the \s-model. In the Lagrangian formulation, this naturally extends to the full 1-form $\J^{\Ah}$ through
\begin{equation}
  \J^{\Ah} = \sEh^{\Ah}{}_I \,\J^I\,, \qquad \text{ with } \qquad \J^{I} = \begin{pmatrix}
    \dd x^i \\
    g_{ij} \star\dd x^j + B_{ij}\, \dd x^j
  \end{pmatrix}\,.
\end{equation}

\subsection{Chiral basis}\label{sec:chiralbasis}

Another basis of $\dlie$, as important as the isotropic one used in the previous subsection, is the so-called chiral basis. Its main motivation is that the defining data of the $\EE$-model is encoded in $\etah_{\Ah\Bh}$, $\Eh_{\Ah\Bh}$ and $\Fh_{\Ah\Bh\Ch}$. However, there is redundancy among these three quantities! In the isotropic basis, we already fixed the form of $\etah_{\Ah\Bh}$. But one can do more and also fix $\Eh_{\Ah\Bh}$ completely, such that all the relevant information about the model is exclusively encoded in the structure coefficients $\Fh_{\Ah\Bh\Ch}$. To this end, we introduce the chiral basis $(\th_{\Ah}) = \bigl( \th_{\ah}, \th_{\ahb} \bigr)$, where $\bigl( \th_{\ah} \bigr)$ and $\bigl( \th_{\ahb} \bigr)$ are orthonormal bases of the eigenspaces $\text{Ker}(\Eh-\mathrm{id})$ and $ \text{Ker}(\Eh + \mathrm{id})$ respectively (recall that $\Eh^2= \mathrm{id}$ and thus that these eigenspaces span the whole algebra $\dlie$). Such a chiral basis is characterised by the simple form taken by $\etah_{\Ah\Bh}$ and $\Eh_{\Ah\Bh}$ in it, namely
\begin{equation}\label{eqn:eta,Echiral}
  \etah_{\Ah\Bh} = \begin{pmatrix}
    \etah_{\,\ah\bh} & 0  \\
    0 & \etah_{\,\ahb\bhb}
  \end{pmatrix} \qquad \text{and} \qquad
  \Eh_{\Ah\Bh} = \begin{pmatrix}
    \etah_{\,\ah\bh} & 0 \\
    0 & - \etah_{\,\ahb\bhb}
  \end{pmatrix}\,,
\end{equation}
where $\etah_{\,\ah\bh} = - \etah_{\,\ahb\bhb}=\text{diag}(+1,\dots,+1,-1,\dots,-1)$ is the flat metric on the tangent space of the target $M=\HH\backslash \DD$, with signature $(p,q)$, $p+q=\text{dim}\,M=\frac{1}{2}\,\text{dim}\, \DD$. The choice of chiral basis is not unique as \eqref{eqn:eta,Echiral} is manifestly invariant under the action of two copies of O($p,q$), forming the so-called double Lorentz group. It relates different choices of orthonormal bases of the eigenspaces $\text{Ker}(\Eh \mp \mathrm{id})$. The \s-model metric and $B$-field are invariant under the action of this double Lorentz group, as can easily be seen from \eqref{eqn:EtogandB}. At the infinitesimal level, such a double Lorentz transformation acts as a change of basis $\delta\th_{\Ah} = U_{\Ah}{}^{\Bh}\,\th_{\Bh}$ with $U_{\ah\bhb}=U_{\,\bhb \ah}=0$ and $U_{\ah\bh}$ and $U_{\ahb\bhb}$ being skew-symmetric matrices.

In particular, the chiral basis results in very simple expressions for the two projectors
\begin{equation}\label{eqn:projectorsPandPb}
  \Ph_{\Ah}{}^{\Bh} =
  \frac12 \left( \delta_{\Ah}{}^{\Bh} + \Eh_{\Ah}{}^{\Bh} \right) =
  \begin{pmatrix}
    \delta_{\ah}{}^{\bh} & 0 \\
    0 & 0
  \end{pmatrix} \qquad \text{and} \qquad
  \Pbh_{\Ah}{}^{\Bh} =
  \frac12 \left( \delta_{\Ah}{}^{\Bh} - \Eh_{\Ah}{}^{\Bh} \right) =
  \begin{pmatrix}
    0 & 0 \\
    0 & \delta_{\ahb}{}^{\bhb}
  \end{pmatrix}
\end{equation}
that can be understood as maps $\Ph: \dlie \rightarrow \text{Ker}(\Eh - \mathrm{id})$ with $\Ph(\th_{\Ah}) = \Ph_{\Ah}{}^{\Bh} \th_{\Bh}$ and similarly for $\Pbh: \dlie \rightarrow \text{Ker}(\Eh + \mathrm{id})$. In the above form, one can see immediately that they satisfy the required properties $\Ph^2 = \Ph$, $\Pbh^2 = \Pbh$ and $\Ph \Pbh = 0$.

\subsection{RG-flow and one-loop \texorpdfstring{$\beta$}{beta}-functions}\label{sec:RG}
Let us finally discuss the renormalisation of $\EE$-models. We fix a double Lie algebra $\dlie = \text{Lie}(\DD)$, with pairing $\langle\cdot,\cdot\rangle$ and a maximally isotropic subalgebra $\hlie=\text{Lie}(\HH)$. The $\EE$-model construction, reviewed above and in Appendix \ref{app:Frame}, then provides a family of $\sigma$-models with target space $M = \HH\backslash \DD$, parametrised by the choice of a symmetric involutive operator $\Eh : \dlie \to \dlie$. Quantum effects will eventually renormalise these $\sigma$-models: in particular, their metric and $B$-field $(g_{ij},B_{ij})$ will acquire a dependence on the Renormalisation Group (RG) scale $\mu$, governed by a differential equation in $\mu$ called the RG-flow. We say that this RG-flow preserves the underlying $\EE$-model structure if the renormalised $\sigma$-models still belong to the family described above. In particular, $\dlie$ and $\hlie$ then remain fixed but there is an induced flow of the operator $\Eh$ with respect to $\mu$. In that case, the Poisson-Lie symmetries and dualities of these $\sigma$-models are preserved under the RG-flow. Moreover, if the expression of the flow of $\Eh$ is independent of the isotropic subalgebra $\hlie$ then Poisson-Lie $T$-dual theories associated to different choices of such isotropic subalgebras will share the same $\beta$-function.

The RG-flow of $\sigma$-models is most often studied through its loop expansion, perturbatively in the quantum parameter $\alpha'$. The fact that this flow preserves the $\EE$-model structure was shown at one-loop in~\cite{Valent:2009nv,Sfetsos:2009dj,Sfetsos:2009vt} and at two-loop in~\cite{Hassler:2020wnp} but currently remains an open question for higher orders. In both cases this flow is independent of $\hlie$. Before describing the one-loop result more explicitly, let us investigate some of the general properties of the RG-flow under the assumption that it preserves the $\EE$-model structure. By construction, such a flow can be written in terms of the $\Eh$-operator as
\begin{equation}\label{eqn:RG}
\frac{\dd \;}{\dd \log \mu} \Eh = \SSh\,,
\end{equation}
where $\SSh$ is a symmetric linear operator on $\dlie$. In this language, Poisson-Lie $T$-dual models share the same $\beta$-function if $\SSh$ depends only on the data $\bigl(\dlie,\langle\cdot,\cdot\rangle,\Eh \bigr)$ and not on the choice of $\hlie$. Moreover, the form of $\SSh$ should be such that the property $\Eh^2=\mathrm{id}$ is preserved along the flow \eqref{eqn:RG}: this requires $\Eh\SSh + \SSh \Eh = 0$.  We will give the explicit expression of $\SSh$ at one-loop later in this subsection.

We now want to translate the abstract equation \eqref{eqn:RG} as an RG-flow for the quantities $\Eh_{\Ah\Bh}$, $\etah_{\Ah\Bh}$ and $\Fh_{\Ah\Bh\Ch}$ used to describe the $\EE$-model in a given choice of basis $( \th_{\Ah} )$ of $\dlie$. This requires a careful treatment, as this basis can in general also depend on the parameters of the theory and could thus itself flow under the RG. A choice often done in the literature is to consider a fixed basis $( \th_{\Ah} )$ in which the structure coefficients $\Fh_{\Ah\Bh\Ch}$ and the pairing $\etah_{\Ah\Bh}$ are constant, which can for instance be an isotropic basis. All the running couplings of the $\EE$-model are then contained in $\Eh_{\Ah\Bh}$ and the equation \eqref{eqn:RG} translates to a flow $\frac{\dd \;}{\dd \log \mu} \Eh_{\Ah\Bh} = \bigl\langle \th_{\Ah}, \SSh\, \th_{\Bh} \bigr\rangle$ on these coefficients. In this choice of basis, the generalised frame field $\sEh^{\Ah}{}_I$ does not flow and the renormalisation of the \s-model couplings $(g_{ij},B_{ij})$ contained in the generalised metric $\Eh_{IJ} = \Eh_{\Ah\Bh} \, \sEh^{\Ah}{}_I \, \sEh^{\Bh}{}_J$ is then only due to the flow of the coefficients $\Eh_{\Ah\Bh}$.

\paragraph{RG-flow in the chiral basis.} In this paper, we will follow a different strategy, by working with the chiral basis $( \th_{\Ah} ) = \bigl( \th_{\ah}, \th_{\ahb} \bigr) $ introduced in the previous subsection. As explained there, this is a basis in which $\etah_{\Ah\Bh}$ and $\Eh_{\Ah\Bh}$ are simultaneously diagonalised in the very simple and constant form \eqref{eqn:eta,Echiral}. We stress that, since the bilinear form $\langle\cdot,\Eh\cdot\rangle$ runs with the RG-flow \eqref{eqn:RG}, the associated chiral basis $( \th_{\Ah} )$ is not constant along the flow. Instead, it evolves through a continuous series of O($D,D$) transformations, which bring back $\langle\cdot,\Eh\cdot\rangle$ to the diagonalised form \eqref{eqn:eta,Echiral} as it runs. We choose to parametrise the flow of the chiral basis as
\begin{equation}\label{eqn:defbetah}
 \frac{\dd}{\dd \log\mu} \left( \th_{\Ah} \right) = \betah_{\Ah}{}^{\Bh}\,\th_{\Bh}\,,
\end{equation}
in terms of a tensor $\betah_{\Ah}{}^{\Bh}$. The fact that the entries $\etah_{\Ah\Bh}$ stay constant along the flow implies that the tensor $\betah_{\Ah\Bh}$ with lowered indices is skew-symmetric, \textit{i.e.} $\betah_{\Ah\Bh} = -\betah_{\Bh\Ah}$. With respect to the decomposition $( \th_{\Ah} ) = \bigl( \th_{\ah}, \th_{\ahb} \bigr)$ of the chiral basis, the components $\betah_{\ah\bh}$ and $\betah_{\,\ahb\bhb}$ generate infinitesimal double Lorentz transformations, corresponding to a redefinition of the orthonormal bases $( \th_{\ah} )$ and $( \th_{\ahb} )$ (see subsection \ref{sec:chiralbasis}), and can thus be set to any skew-symmetric values without affecting the flow of the underlying \s-model. All the physical information about the RG-flow is then contained in the components $\betah_{\ah\bhb} = -\betah_{\,\bhb\ah}$. These can be related to the operator $\SSh$ characterising the flow \eqref{eqn:RG} of $\Eh$ by imposing that, by definition of the chiral basis, the entries $\Eh_{\Ah\Bh} = \langle \th_{\Ah}, \Eh\,\th_{\Bh} \rangle$ keep the diagonalised form \eqref{eqn:eta,Echiral} along the flow. We then find that
\begin{equation}
  \betah_{\ah\bhb} = \frac12 \langle \th_{\ah}, \SSh\, \th_{\,\bhb} \rangle
    \qquad\text{and}\qquad
  \betah_{\,\ahb\bh} = -\frac12 \langle \th_{\ahb}, \SSh\, \th_{\bh} \rangle\,.
\end{equation}
Let us also note that the preservation of $\Eh^2 = \mathrm{id}$ along the flow imposes $\langle \th_{\ah}, \SSh\, \th_{\bh} \rangle = \langle \th_{\ahb}, \SSh\, \th_{\,\bhb} \rangle = 0$. As expected, the data of the operator $\SSh$ is thus equivalent to that of the components $\betah_{\ah\bhb}$. This observation originates from the two different ways of encoding the RG-flow: the former in terms of the flow of $\Eh$ and the latter in terms of the flow of the chiral basis.

It is clear that the generalised frame field coefficients $\sEh^{\Ah}{}_I$ in the chiral basis are not RG-invariant, since the basis $(\th_{\Ah})$ itself runs with $\mu$. The flow of $\sEh^{\Ah}{}_I$ is simply given by
\begin{equation}\label{eqn:flowFrame}
\frac{\dd}{\dd \log\mu} \left( \sEh^{\Ah}{}_I \right) = \betah^{\Ah}{}_{\Bh} \, \sEh^{\Bh}{}_I\,,
\end{equation}
providing a useful geometric interpretation of the tensor $\betah^{\Ah}{}_{\Bh}$.
In that framework, the renormalisation of the \s-model couplings $(g_{ij},B_{ij})$ contained in the generalised metric $\Eh_{IJ} = \Eh_{\Ah\Bh} \, \sEh^{\Ah}{}_I \, \sEh^{\Bh}{}_J$ is then only due to the flow of the frames $\sEh^{\Ah}{}_I$, since the coefficients $\Eh_{\Ah\Bh}$ are kept constant.

Let us finally recall that in the chiral basis, all the parameters of the $\EE$-model are encoded in the structure coefficients $\Fh_{\Ah\Bh\Ch}$. The RG-flow of the theory can then also be conveniently written in terms of these coefficients and the tensor $\betah_{\Ah}{}^{\Bh}$, taking the form
\begin{equation}\label{eqn:betaforFh}
  \frac{\dd \Fh_{\Ah\Bh\Ch}}{\dd \log\mu} = 3 \betah_{[\Ah}{}^{\Dh} \Fh_{\Bh\Ch]\Dh}\,.\vspace{4pt}
\end{equation}

\paragraph{One-loop $\beta$-functions.} We end this subsection by the explicit description of the RG-flow of $\EE$-models at one-loop~\cite{Valent:2009nv,Sfetsos:2009dj,Sfetsos:2009vt}. Following the general discussion above, this flow can be encoded in the operator $\SSh$ or equivalently in the tensor $\betah_{\ah\bhb} = \frac12 \langle \th_{\ah}, \SSh \,\th_{\,\bhb} \rangle$. At one-loop, \textit{i.e.} at the first order in the $\alpha'$-expansion, the latter takes the particularly simple form~\cite{Hassler:2020wnp}
\begin{equation}\label{eqn:beta-1loop}
\betah_{\ah\bhb}^{(1)} = - \Fh_{\,\ah\chb}{}^{\dh}\, \Fh_{\,\bhb\dh}{}^{\chb}\,.
\end{equation}

\section{Affine Gaudin models and integrability}\label{sec:integrability}
We are now specialising to a class of integrable $\EE$-models called affine Gaudin models \cite{Vicedo:2017cge,Lacroix:2020flf}. At their heart is a Lax connection, which is fixed by a rational function $\varphi(z)$ of the spectral parameter $z$, the twist function.

\subsection{Lax connection and Maillet bracket}
In order to identify a field theory as integrable, we have to construct an infinite number of independent conserved charges. Moreover, these charges are required to be in involution, meaning that they should Poisson commute. A systematic approach to this problem relies on identifying a Lax connection $\L(z) = \L_\sigma(z) \dd \sigma + \L_\tau(z) \dd \tau$ built out of the fields of the theory. This is a one-parameter family of one-forms on the worldsheet, valued in (the complexification of) a simple Lie algebra $\fg$ and labelled by the spectral parameter $z\in\mathbb{C}$, which is required to be on-shell flat, namely such that
\begin{equation}\label{eqn:flatLax}
  \dd \L(z) + \L(z) \wedge \L(z) = 0 \,,
\end{equation}
for all values of $z$, after imposing the field equations. Integrating the Lax connection $\L(z)$ along a spatial path $\sigma \in [0, 2\pi]$, we obtain the monodromy\footnote{Because $\L_\sigma$ is Lie algebra valued and therefore non-commutative, we have to use the path-ordered exponential.}
\begin{equation}
  T(z) = P \overleftarrow{\exp} \left( - \int_0^{2\pi} \L_\sigma(z) \dd \sigma \right)\,.
\end{equation}
Combining \eqref{eqn:flatLax} with periodic boundary conditions $\L(\sigma$=$0;z)=\L(\sigma$=$2\pi;z)$, one finds that the time evolution of $T(z)$ is governed by 
\begin{equation}
  \partial_\tau T(z) = \left[\L_\tau(0;z),T(z)\right]\,.
\end{equation}
It follows that the trace $\tr(T(z))$ is conserved for all $z$, even though $T(z)$ itself is in general not conserved. Another common setup in the literature assumes boundary conditions where the Lax connection vanishes at $\sigma=\pm\infty$. In this case $T(z)$ is conserved.

An infinite number of conserved charges can be obtained by expanding $\tr(T(z))$ around a specific point. A sufficient condition ensuring the involution of these conserved charges is that the Poisson brackets of the Lax connection's spatial component are of Maillet's $r$/$s$ form \cite{Maillet:1985ek, Maillet:1985ec}
\begin{align}\label{eqn:rs}
\{\L_\sigma^\alpha(\sigma_1;z), \L_\sigma^\beta(\sigma_2;w)\} &= 
(r+s)^{\delta\beta}(z,w) f_{\delta\gamma}{}^\alpha \L_\sigma^\gamma(\sigma_1;z) \delta(\sigma_1-\sigma_2) \\ 
&+ (r-s)^{\alpha\delta}(z,w) f_{\delta\gamma}{}^\beta \L_\sigma^\gamma(\sigma_2;w) \delta(\sigma_1-\sigma_2)
- 2 s^{\alpha\beta}(z,w) \partial_{\sigma_1}\delta(\sigma_1-\sigma_2)\,,\notag
\end{align}
for some functions $r^{\alpha\beta}(z,w)$ and $s^{\alpha\beta}(z,w)$.
Here, Greek indices like $\alpha$, $\beta$, $\ldots$ label the generators $t_\alpha$ of the Lie algebra $\fg$ in which the Lax connection takes value. Moreover, we have the corresponding structure constants, defined by 
\begin{equation}
  [t_\alpha, t_\beta] = f_{\alpha\beta}{}^\gamma t_\gamma\,,
\end{equation}
and $\L_\sigma^\alpha$ are the coefficients of the spatial component of the Lax connection $\L_\sigma = t_\alpha \L_\sigma^\alpha$. 

For the class of integrable models we are interested in throughout the remainder of this article, one can express $r$ and $s$ in terms of an invariant metric $\kappa^{\alpha\beta}$ on $\fg$ and a rational function $\varphi(z)$, called the twist function, as
\begin{equation}
  (r+s)^{\alpha\beta}(z,w) = \frac{\varphi(w)^{-1}}{w-z} \kappa^{\alpha\beta}\,, \quad
  (r-s)^{\alpha\beta}(z,w) = \frac{\varphi(z)^{-1}}{w-z} \kappa^{\alpha\beta}\,.
\end{equation}
Our next goal is to explain how the analysis of a Maillet bracket with twist function naturally leads to the systematic construction of integrable $\mathcal{E}$-models. By doing so, we reveal the relation between these integrable $\mathcal{E}$-models and the formalism of affine Gaudin models. Finally, we will discuss two equivalent but complementary descriptions, each presenting their respective benefits, and corresponding to different bases of the underlying Lie algebra, which are naturally associated to either the poles or the zeroes of the twist function $\varphi(z)$.

\subsection{Affine Gaudin models and poles basis}\label{sec:poles}
The deeper algebraic origin behind the non-ultralocal Poisson algebra \eqref{eqn:rs} is best understood by working with the combination $\Gamma^\alpha(\sigma; z) = \varphi(z) \L_\sigma^\alpha(\sigma; z)$ which satisfies the Poisson bracket
\begin{align}\label{eqn:rs Gaudin}
  \{\Gamma^\alpha(\sigma_1;z), \Gamma^\beta(\sigma_2;w)\} &= 
    - f^{\alpha\beta}{}_\gamma \frac{\Gamma^\gamma(\sigma_1;z) -\Gamma^\gamma(\sigma_2;w)}{z-w} \delta(\sigma_1-\sigma_2) \notag\\
  &\qquad\qquad + \frac{\varphi(z) - \varphi(w)}{z-w} \kappa^{\alpha\beta} \partial_{\sigma_1}\delta(\sigma_1-\sigma_2)\,.
\end{align}
This simple rewriting of \eqref{eqn:rs} leads to the interpretation of the $\fg$-valued connection $\varphi(z) \partial_\sigma + t_\alpha \Gamma^\alpha(\sigma; z)$ as the Lax matrix of an affine Gaudin model \cite{Levin:2001nm,Feigin:2007mr,Vicedo:2017cge}, see also \cite{Delduc:2019bcl}. If $\varphi(z)$ has poles at $z_i$ for $i = 1, \ldots, M$ of orders $n_i \geq 1$ then it is consistent with \eqref{eqn:rs Gaudin} to take $\Gamma^\alpha$ to have the same pole structure, so that
\begin{equation}\label{eqn:GaudinLax}
  \varphi(z) = \sum_{i=1}^{M} \sum_{p=0}^{n_i-1} \frac{\ell_{[i, p]}}{(z - z_i)^{p+1}} + \lambda^{-1}, \qquad
  \Gamma^\alpha(\sigma;z) = \sum_{i=1}^{M} \sum_{p=0}^{n_i-1} \frac{J_\sigma^{\alpha [i, p]}(\sigma)}{(z - z_i)^{p+1}}\,
\end{equation}
for some constant parameters $\ell_{[i,p]}$ and $\lambda$ and some fields $J_\sigma^{\alpha [i,p]}(\sigma)$. It then immediately follows from the bracket \eqref{eqn:rs Gaudin} that the coefficients of the Gaudin Lax matrix \eqref{eqn:GaudinLax} satisfy the current algebra
\begin{align}\label{eqn:PBJ Gaudin}
  \bigl\{J_\sigma^{\alpha \widetilde A}(\sigma_1), J_\sigma^{\beta \widetilde B}(\sigma_2) \bigr\} &= f^{\alpha\beta}{}_\gamma \,
  \widetilde F^{\widetilde A\widetilde B}{}_{\widetilde C}\, J_\sigma^{\gamma \widetilde C}(\sigma_1)\, \delta(\sigma_1-\sigma_2) - \kappa^{\alpha\beta}\, \widetilde\eta^{\widetilde A\widetilde B}\, \partial_{\sigma_1}\delta(\sigma_1-\sigma_2)
\end{align}
where we use the single index notation $\widetilde A$ for a pair of indices $[i, p]$ where $i = 1, \ldots, M$ labels the sites of the affine Gaudin model (\textit{i.e.} the poles of $\varphi(z)$) and $p = 0, \ldots, n_i - 1$ the corresponding order of the poles.
Explicitly, we have
\begin{equation} \label{tildeF}
  \widetilde F^{\widetilde A\widetilde B}{}_{\widetilde C} = \begin{cases}
    1 & \text{if } \widetilde A=[i,p], \widetilde B=[i,q] \text{ and } \widetilde C=[i, p+q]\\
    0 & \text{otherwise}
  \end{cases}
\end{equation}
and
\begin{equation}\label{eqn:tildeEta}
  \widetilde \eta^{\widetilde A\widetilde B} = \begin{cases}
    \ell_{[i, p+q]} & \text{if } \widetilde A=[i,p] \text{ and } \widetilde B=[i,q]\\
    0 & \text{otherwise}.
  \end{cases}
\end{equation}
In other words, \eqref{eqn:PBJ Gaudin} describes a current bracket with underlying Lie algebra given by a direct sum of truncated loop algebras, or Takiff algebras, over $\fg$ and with levels given by the $\ell_{\widetilde A}$.

So far we have shown that the Maillet bracket \eqref{eqn:rs} with twist function $\varphi(z)$ is equivalent to the data of currents $J_\sigma^{\alpha \widetilde A}(\sigma)$ associated with the poles of $\varphi(z)$ and satisfying the simple Poisson bracket \eqref{eqn:PBJ Gaudin}. The remaining ingredient needed to construct an integrable field theory is the choice of an Hamiltonian $H$ such that the dynamics $\partial_\tau = \lbrace H, \cdot \rbrace$ takes the form of a flatness equation \eqref{eqn:flatLax} for the Lax matrix $\L_\sigma(z)$. This is exactly what the affine Gaudin model formalism provides. More precisely, it was shown in~\cite{Vicedo:2017cge,Delduc:2019bcl} that the time evolution of $\L_\sigma(z)$ takes the form of a flatness equation if one chooses the Hamiltonian to be
\begin{equation}\label{eqn:HGamma}
H = - \sum_{A=1}^{2N} \res_{z=\zeta_A} \frac{\epsilon_A}{2\varphi(z)} \int_0^{2\pi} \dd\sigma\,\kappa_{\alpha\beta}\, \Gamma^\alpha(\sigma,z) \Gamma^\beta(\sigma,z),
\end{equation}
where the points $\lbrace \zeta_A \rbrace_{A=1,\dots,2N}$ are defined as the zeroes\footnote{The number of zeroes of the twist function $\varphi(z)$ in equation \eqref{eqn:GaudinLax} is given by $\sum_{i=1}^M n_i$, where $n_i$ is the multiplicity of the pole $z_i$. We will suppose here that this number is even and will denote it as $2N$.} of $\varphi(z)$, which we suppose real and simple, and the coefficients $\epsilon_A$ are constant numbers. As explained in~\cite{Delduc:2019bcl}, to ensure that the model is relativistic and has a positive Hamiltonian when $\fg$ is compact, one further has to take $\epsilon_A = -\text{sign}\bigl(\varphi'(\zeta_A)\bigr)$. For reasons to be explained later, we suppose from now on that the choice of $\varphi(z)$ is such that there are as many $\epsilon_A$'s equal to $+1$ as $\epsilon_A$'s equal to $-1$. Reinserting the expression \eqref{eqn:GaudinLax} of $\Gamma(\sigma,z)$ in the above equation, one can compute explicitly the residues and write the Hamiltonian in the form
\begin{equation}\label{eqn:HGaudin}
H = \frac12\int_0^{2\pi} \dd \sigma \, \kappa_{\alpha\beta}\, \widetilde{\mathcal{E}}_{\widetilde A\widetilde B}\, J_\sigma^{\alpha \widetilde A} J_\sigma^{\beta \widetilde B},
\end{equation}
where the explicit expression of the coefficients $\widetilde{\mathcal{E}}_{\widetilde A\widetilde B}$, which depend on the location of the zeroes $\zeta_A$, the poles $z_i$ and their orders, as well as the coefficients $\epsilon_A$, will not be needed in what follows.

The above results are very suggestive of a relation between affine Gaudin models and $\mathcal{E}$-models. Indeed, the Poisson bracket \eqref{eqn:PBJ Gaudin} of the currents $J_\sigma^{\alpha \widetilde A}(\sigma)$ obtained here from the Maillet bracket is a special case of the fundamental Poisson algebra \eqref{eqn:PBJh} defining an $\mathcal{E}$-model. More precisely, this $\mathcal{E}$-model is built from a Lie algebra $\mathfrak{d}$ with a basis labelled by pairs of indices $\alpha\widetilde A$, where the $\alpha$'s label a basis of $\fg$ and the $\widetilde A$'s are indices associated with the poles of $\varphi(z)$, as introduced earlier in this subsection. Using these indices, the structure constants of this algebra $\mathfrak{d}$ then take a factorised form $f^{\alpha\beta}{}_\gamma\,\widetilde F^{\widetilde A\widetilde B}{}_{\widetilde C}$, where $f^{\alpha\beta}{}_\gamma$ are the structure constants of $\fg$ and $\widetilde F^{\widetilde A\widetilde B}{}_{\widetilde C}$ are defined in equation \eqref{tildeF}. This identifies $\mathfrak{d}$ with the direct sum of Takiff algebras of multiplicity $n_i$ associated with the poles of $\varphi(z)$.

Similarly, the invariant pairing on $\mathfrak{d}$ that arises from the above analysis takes the factorised form $\kappa^{\alpha\beta}\,\widetilde \eta^{\widetilde A\widetilde B}$, where $\widetilde \eta^{\widetilde A\widetilde B}$ is defined in equation \eqref{eqn:tildeEta} in terms of the coefficients in the partial fraction decomposition of $\varphi(z)$. Finally, the operator $\Eh$ characterising the $\mathcal{E}$-model is obtained by comparing the expressions \eqref{eqn:EmodelH} and \eqref{eqn:HGaudin} for the Hamiltonian of the theory. The matrix entries of this operator then take a factorised form $\delta^{\beta}{}_\alpha\, \widetilde{\mathcal{E}}^{\widetilde B}{}_{\widetilde A}$ with $\widetilde{\mathcal{E}}^{\widetilde B}{}_{\widetilde A}$ defined through equation \eqref{eqn:HGaudin}. One can show that this operator is symmetric with respect to the pairing defined above and is an involution with $\Eh^2 = \text{id}$, using the fact that the coefficients $\epsilon_A$ square to $1$ (indeed, these coefficients turn out to be the eigenvalues of $\Eh$).

The various ingredients described above define an $\mathcal{E}$-model, which by construction is integrable. This thus shows that any integrable field theory with twist function $\varphi(z)$ and Lax matrix $\mathcal{L}_\sigma(z) = \varphi(z)^{-1} \Gamma(z)$ given by \eqref{eqn:GaudinLax} can be naturally cast as an $\mathcal{E}$-model. These theories exactly coincide with the models introduced in~\cite{Lacroix:2020flf} using the 4-dimensional Chern-Simons theory, whose Hamiltonian formulation is known to be deeply related to the affine Gaudin model construction~\cite{Vicedo:2019dej}. In this subsection, we have described the Lie algebra $\mathfrak{d}$, the invariant pairing $\langle\cdot,\cdot\rangle$ and the operator $\Eh$ defining the $\mathcal{E}$-model in a basis naturally associated with the poles of the twist function $\varphi(z)$. In this description, the structure constants of the $\mathcal{E}$-model are particularly simple -- see the definition \eqref{tildeF} of $\widetilde F^{\widetilde A\widetilde B}{}_{\widetilde C}$, allowing an easy identification of the underlying Lie algebra $\mathfrak{d}$. This is for instance useful to construct the maximally isotropic subalgebra $\mathfrak{h}$ of $\mathfrak{d}$ which is needed to relate the $\mathcal{E}$-model to a standard $\sigma$-model. However, the price to pay for this simplicity is a more complicated expression for the invariant pairing -- see the definition \eqref{eqn:tildeEta} of $\widetilde \eta^{\widetilde A\widetilde B}$ -- and a quite convoluted expression for the operator $\Eh$.

\subsection{Zeroes basis}\label{sec:zeroes}
In the above affine Gaudin model formulation of integrable $\mathcal{E}$-models, all the couplings of the theory are encoded in the tensors $\widetilde \eta^{\widetilde A \widetilde B}$ and $\widetilde{\mathcal{E}}_{\widetilde A\widetilde B}$ since the structure constants \eqref{tildeF} are independent of the couplings. It was argued in \cite{Hassler:2020wnp} that for the purpose of computing the one- and two-loop RG flows of these $\sigma$-models, it is convenient to encode all the $\mathcal{E}$-model couplings in the structure coefficients themselves: we refer to subsections \ref{sec:chiralbasis} and \ref{sec:RG} for a review. In particular, this approach was used in \cite{Hassler:2020xyj} to compute the $1$-loop RG flow of the integrable $\mathcal E$-models under consideration in the case when $\varphi(z)$ has at most simple poles.

The change of basis in the current algebra \eqref{eqn:PBJ Gaudin} which has the effect of both diagonalising the tensors $\widetilde \eta_{\widetilde A \widetilde B}$ and $\widetilde{\mathcal{E}}_{\widetilde A\widetilde B}$ and removing all of their coupling dependence was constructed in \cite{Lacroix:2020flf}. It can be motivated by analysing the definition \eqref{eqn:HGamma} of the Hamiltonian of these models. Indeed, this expression shows that the Hamiltonian is more naturally expressed in terms of the zeroes $\lbrace \zeta_A \rbrace_{A=1,\dots,2N}$ of the twist function, rather than its poles. Because the Hamiltonian of an $\mathcal{E}$-model is directly related to its operator $\Eh$, this suggests that the latter takes a simpler form in a basis associated with these zeroes. One can obtain this alternative form of the model from the results of the previous subsection by performing explicitly the change of basis suggested in \cite{Lacroix:2020flf}, in particular to obtain the expression of the structure constants in this new basis. For completeness, we present this computation in Appendix \ref{app:ChangeBasis}. Here, we will follow an alternative route and rederive the form of these integrable $\mathcal{E}$-models in the basis associated with zeroes directly from the Maillet bracket, to show that it naturally arises from the integrable structure of the theory. As expected, the two procedures lead to the same results, as can be checked by comparison with Appendix \ref{app:ChangeBasis}.

Since we are now working with the zeroes of the twist function rather than its poles, it is natural to consider the partial fraction decomposition of its inverse. The latter is given by
\begin{equation}\label{eqn:phiinv}
\varphi(z)^{-1} = -\sum_{A=1}^{2 N} \frac{\epsilon_A L_A^2}{z-\zeta_A} + \lambda\,,
\end{equation}
where we introduce
\begin{equation}
L_A = |\varphi'(\zeta_A)|^{-1/2}
\end{equation}
and where $\epsilon_A = -\text{sign}\bigl(\varphi'(\zeta_A)\bigr)$ as in the previous subsection. We now define currents $\J_\sigma^{\alpha A}(\sigma)$ associated with the zeroes $\zeta_A$ of $\varphi(z)$ by considering the following ansatz for the partial fraction decomposition of the Lax matrix:
\begin{equation}\label{eqn:ourLax}
\L_\sigma^\alpha(\sigma;z) = \sum_{A=1}^{2N} \frac{L_A \J_\sigma^{\alpha A}(\sigma)}{z - \zeta_A}\, ,
\end{equation}
where we introduced the normalisation by $L_A$ for future convenience. It now follows that the Maillet bracket \eqref{eqn:rs} is equivalent to the current algebra
\begin{align}\label{eqn:PBJ}
\bigl\{\J_\sigma^{\alpha A}(\sigma_1), \J_\sigma^{\beta B}(\sigma_2) \bigr\} &= f^{\alpha\beta}{}_\gamma \,
F^{AB}{}_C \,\J_\sigma^{\gamma C}(\sigma_1)\, \delta(\sigma_1-\sigma_2) - \kappa^{\alpha\beta}\, \eta^{AB} \,\partial_{\sigma_1}\delta(\sigma_1-\sigma_2)\,,
\end{align}
where the tensor $\eta^{AB}$ is defined as
\begin{equation}\label{eqn:eta}
  \eta^{AB} = \begin{cases}
    \epsilon_A & \text{if }A=B \\
    0 & \text{otherwise}
  \end{cases}\,,
\end{equation}
while the totally symmetric tensor $F_{ABC}$ is zero if all indices are different and the remaining non-zero components are given by
\begin{subequations}\label{eqn:genfluxes}
\begin{align}
F_{AAB} &= \frac{\epsilon_A L_B}{\zeta_A - \zeta_B} \quad \text{and} \\
F_{AAA} &= \frac{1}{L_A} \left( \lambda - \sum_{B\ne A}^{2 N} \frac{\epsilon_B L_B^2}{\zeta_A - \zeta_B} \right) \, .
\end{align}
\end{subequations}
Moreover, in this basis the Hamiltonian \eqref{eqn:HGamma} takes the simple form
\begin{equation}
  H = \frac12\int_0^{2\pi} \dd \sigma \, \kappa_{\alpha\beta}\, \mathcal{E}_{AB}\, \J_\sigma^{\alpha A} \J_\sigma^{\beta B} \quad\text{with}\quad
  \mathcal{E}_{AB}=\begin{cases}
    1 & \text{if }A=B\\
    0 & \text{otherwise}\,.
  \end{cases}
\end{equation}

We now have all the necessary ingredients to interpret this integrable field theory as an $\mathcal{E}$-model. Following the notations of section \ref{sec:PLsym}, the Lie algebra $\mathfrak{d}$ underlying this model has a basis $\lbrace \th_{\Ah} \rbrace$ which in the present case is labelled by pairs of indices $\Ah = \alpha \, A$, where $\alpha$ labels the basis $\lbrace t_\alpha \rbrace$ of $\fg$ and $A$ labels the zeroes $\zeta_A$. The structure constants $\Fh_{\Ah\Bh\Ch}$ and invariant pairing $\etah_{\Ah\Bh}$ of $\mathfrak{d}$ are defined through the factorised expressions:
\begin{equation}\label{eqn:hattonohat1}
  \Fh_{\Ah\Bh\Ch} = f_{\alpha\beta\gamma}\, F_{ABC} \qquad \text{ and } \qquad \etah_{\Ah\Bh} = \kappa_{\alpha\beta}\, \eta_{AB},
\end{equation}
with $F_{ABC}$ and $\eta_{AB}$ as above. Finally, the operator $\Eh$ is defined by
\begin{equation}\label{eqn:hattonohat2}
  \Eh ( \th_{\Ah} ) = \th_{\Bh}\, \Eh^{\Bh}{}_{\Ah}, \qquad \text{ with } \qquad \Eh^{\Bh}{}_{\Ah} = \delta^{\beta}{}_{\alpha}\,\delta^B{}_A\,\epsilon_A.
\end{equation}
As expected, in this basis associated with the zeroes of the twist function, the pairing $\langle\cdot,\cdot\rangle$ and the operator $\Eh$ are extremely simple, and in particular independent of the parameters of the model. The latter are then completely encoded in the structure constants, which take a more complicated form -- see equation \eqref{eqn:genfluxes}, making the structure of the underlying Lie algebra $\dlie$ less transparent. However, we know from the results of the previous subsection (and more explicitly the appendix \ref{app:ChangeBasis}) that $\mathfrak{d}$ can be identified with a direct sum of Takiff algebras: indeed, in the other basis associated with the poles of the twist function, the structure constants become much simpler, at the cost of a more complicated expression for the entries of the bilinear form and the operator $\Eh$.

Using the terminology of subsection \ref{sec:chiralbasis}, the zeroes basis considered here is identified with the chiral basis of the integrable $\EE$-model, as we now explain. First note that we are free to choose the basis $\lbrace t_\alpha \rbrace_{\alpha=1, \ldots, \text{dim}\,\fg}$ of $\fg$ to be orthonormal, such that the bilinear form $\kappa_{\alpha\beta} = \text{diag}(+1,\dots,+1,-1,\dots,-1)$ is diagonalised with entries normalised to $+1$ or $-1$. Next, recall that we supposed earlier that there are as many coefficients $\epsilon_A$ equal to $+1$ than there are equal to $-1$. Because these coefficients are the entries of the diagonal matrix $\eta_{AB}$ in equation \eqref{eqn:eta}, this fact ensures that the bilinear form $\etah_{\Ah\Bh}=\kappa_{\alpha\beta}\,\eta_{AB}$ is of split signature, as required in the definition of an $\EE$-model. We use this observation to separate the indices $A$ into two groups of equal size: $a\in\lbrace 1,\dots,N\rbrace$ with $\epsilon_a=+1$ and $\ab \in \lbrace 1,\dots,N\rbrace$ with $\epsilon_{\ab}=-1$. The zeroes basis $\lbrace \th_{\Ah} = \th_{\alpha\, A} \rbrace$ of $\dlie$ then naturally decomposes into two subsets $\lbrace \th_{\ah} = \th_{\alpha\,a} \rbrace_{\alpha=1,\dots,\text{dim}\,\fg}^{a=1,\dots,N}$ and $\lbrace \th_{\ahb} = \th_{\alpha\,\ab} \rbrace_{\alpha=1,\dots,\text{dim}\,\fg}^{\ab=1,\dots,N}$  of size $N\,\text{dim}\,\fg$ each. In this basis $\etah_{\Ah\Bh}$ and $\Eh_{\Ah\Bh}$ take the diagonalised form \eqref{eqn:eta,Echiral}, thus identifying the zero basis with the chiral one. As argued in subsection \ref{sec:RG}, the latter is very well adapted for the study of the RG-flow of the theory, which will be the subject of the next section.

\section{Renormalisation and 1-loop \texorpdfstring{$\beta$}{beta}-functions\label{sec:betafuncs}}
Quantum effects will eventually renormalise the action \eqref{eqn:smodel}. Consequently, we have to check if the resulting effective action is still governed by the integrable structure introduced above. As explained in subsections \ref{sec:chiralbasis} and \ref{sec:RG}, for a general $\EE$-model in the chiral basis, these quantum effects will renormalise the defining data of the theory, namely its structure constants $\Fh_{\Ah\Bh\Ch}$ (recall that in the chiral basis, the coefficients $\etah_{\Ah\Bh}$ and $\Eh_{\Ah\Bh}$ are fixed and do not contain any parameters of the theory). For the integrable $\EE$-models under investigation, we have identified the chiral basis as the one associated with the zeroes of the twist function $\varphi(z)$ in subsection \ref{sec:zeroes}. In this basis, the structure constants of the theory are expressed in terms of particular combinations \eqref{eqn:genfluxes} of the parameters $\lambda$, $\zeta_A$ and $L_A$ defining $\varphi(z)$. A crucial ingredient in addressing the question of the renormalisability of the models under consideration is to understand the map between infinitesimal deformations of the twist function and the corresponding deformations of the structure constants. This is the subject of the first subsection.

\subsection{Variations of the twist function and structure coefficients}\label{sec:var}
Let us start with a brief counting of parameters. According to equation \eqref{eqn:phiinv}, $\varphi(z)$ encodes $4 N + 1$ parameters $\lambda$, $\zeta_A$ and $L_A$, which we refer to as affine Gaudin parameters and whose space we denote by $\Pi$. However, we note that the structure coefficients \eqref{eqn:genfluxes} only contain differences $\zeta_A-\zeta_B$. Thus, shifting all $\zeta_A$'s by a constant leaves them invariant. Similarly, homogeneously scaling $\lambda$, $\zeta_A$ and $L_A$ by a constant also has no effect on the structure coefficients. This translates the invariance of the integrable $\EE$-model under modifications of the twist function $\varphi(z)$ which can be reabsorbed via translations and dilations of the spectral parameter. Without loss of generality, these operations can be used to eliminate 2 of the parameters among $\lambda$, $\zeta_A$ and $L_A$ (for instance by setting $\lambda=1$ and $\sum_{A=1}^{2N} \zeta_A = 0$), implying that there are only $4 N - 1$ independent physical parameters. In other words, the space of parameters of the model is a $(4 N - 1)$--dimensional quotient of $\Pi$.

\paragraph{Variation of the twist function.} Let us now consider infinitesimal variations $\dd\lambda$, $\dd\zeta_A$ and $\dd L_A$ of the affine Gaudin parameters, which we can see as one-forms on $\Pi$. This induces a variation $\dd \varphi(z)$ of the twist function. It is always possible, and will be convenient to reparametrise, this variation as
\begin{equation}\label{eqn:dphi}
  \dd \varphi(z) = e(z) \varphi(z) + \partial_z \left[ f(z) \varphi(z) \right]\,
\end{equation}
in terms of two functions
\begin{equation}
  e(z) = \sum_{A=1}^{2 N} \frac{\epsilon_A L_A^2 e_A}{z - \zeta_A}
    \quad\text{and}\quad
    f(z) = \sum_{A=1}^{2 N} \frac{\epsilon_A L_A^2 f_A}{z - \zeta_A} + f_t + f_d\,z\,,
\end{equation}
where $e_A$, $f_A$, $f_t$ and $f_d$ are $4 N + 2$ one-forms on $\Pi$. These one-forms can be related to the fundamental variations $\dd L_A$, $\dd \zeta_A$ and $\dd\lambda$ by matching the right- and left-hand side of
\begin{equation}
  \dd \big( \varphi(z)^{-1} \big) = -e(z) \varphi(z)^{-1} - \partial_z f(z) \varphi(z)^{-1} + f(z) \partial_z \big( \varphi(z)^{-1} \big)
\end{equation}
yielding
\begin{subequations}\label{eqn:dGaudin}
\begin{align}
  \dd L_A &= - L_A \left[ f_d - \frac{ \lambda e_A }2 + \sum_{B\ne A}^{2 N} \frac{\epsilon_B L_B^2}{\zeta_A - \zeta_B}\left( \frac{f_A - f_B}{\zeta_A - \zeta_B} +
    \frac{e_A + e_B}2  \right) \right] \\
  \dd \zeta_A &= - f_t - f_d\,\zeta_A - f_A\,\lambda - \epsilon_A L_A^2 e_A + \sum_{B\ne A}^{2N} \frac{\epsilon_B L_B^2 (f_A - f_B)}{\zeta_A - \zeta_B} \\
  \dd \lambda &= -f_d \lambda \,.
\end{align}
\end{subequations}
As expected from a simple counting argument, the relation $(e_A,f_A,f_t,f_d) \mapsto (\dd L_A, \dd \zeta_A, \dd\lambda)$ is not injective. Indeed, the above expressions for $(\dd L_A, \dd \zeta_A, \dd\lambda)$ are invariant under a shift of $(e_A,f_A,f_t,f_d)$ by $(0,-c,\lambda\,c,0)$, where $c$ is an arbitrary one-form. In terms of the functions $e(z)$ and $f(z)$ considered above, this corresponds to a shift of $f(z)$ by $c\,\varphi(z)^{-1}$, as can be seen from the equation \eqref{eqn:phiinv}. It is clear that such a shift does not modify the variation \eqref{eqn:dphi} of $\dd\varphi(z)$ and thus encodes a redundancy in its parametrisation by $e(z)$ and $f(z)$. On the other hand, one shows that the map $(e_A,f_A,f_t,f_d) \mapsto (\dd L_A, \dd \zeta_A, \dd\lambda)$ is surjective. Therefore, we can invert the above relation and find the expression of the one-forms $(e_A,f_A,f_t,f_d)$ in terms of $(\dd L_A, \dd \zeta_A, \dd\lambda)$, up to an arbitrary shift by $(0,-c,\lambda\,c,0)$. We will not need the explicit form of this expression and thus skip it for simplicity. To summarise, any infinitesimal variation $(\dd L_A, \dd \zeta_A, \dd\lambda)$ of the affine Gaudin parameters can be parametrised as in \eqref{eqn:dphi} in terms of the functions $e(z)$ and $f(z)$. This parametrisation is unique up to the shifts $f(z) \mapsto f(z) + c\,\varphi(z)^{-1}$.

\paragraph{Variation of the structure coefficients.} Let us finally investigate the corresponding infinitesimal transformation of the structure coefficients. Recall that they take a factorised form $\Fh_{\Ah\Bh\Ch} = f_{\alpha\beta\gamma} \, F_{ABC}$, where the first factor depends only on the Lie algebra structure of $\fg$ and the second factor is the combination \eqref{eqn:genfluxes} of the affine Gaudin parameters. Under a variation of the latter, these structure coefficients transform as
\begin{equation}\label{eqn:dF=TF}
  \dd \Fh_{\Ah\Bh\Ch} = 3 \Th_{[\Ah}{}^{\Dh} \Fh_{\Bh\Ch]\Dh}
\end{equation}
where
\begin{equation}\label{eqn:defT}
  \Th_{\Ah}{}^{\Bh} = \delta_\alpha{}^\beta T_A{}^B
    \quad\text{and}\quad
  \dd F_{ABC} = 3 T_{(A}{}^D F_{BC)D}\,.
\end{equation}
In terms of the parametrisation $(e_A,f_A,f_t,f_d)$, we find that $T_{AB} = (T_\mathrm{a})_{AB} + (T_\mathrm{s})_{AB}$ decomposes into the antisymmetric part
\begin{subequations}\label{eqn:T}
\begin{equation}\label{eqn:Ta}
  (T_\mathrm{a})_{AB} = \begin{cases}\displaystyle
    - \frac{L_A L_B}{\zeta_A - \zeta_B}\left( \frac{f_A - f_B}{\zeta_A - \zeta_B} +
      \frac{e_A + e_B}{2}
    \right) & A\ne B \\[0.25em]\displaystyle
    0 & A = B
  \end{cases}
\end{equation}
and the symmetric part
\begin{equation}\label{eqn:Ts}
  (T_\mathrm{s})_{AB} = \begin{cases}\displaystyle
    - L_A L_B \frac{e_A - e_B}{6 (\zeta_A - \zeta_B)} & A \ne B \\[0.25em]\displaystyle
    \frac{\epsilon_A}6 \left( \sum_{C\ne A}^{2N} \frac{\epsilon_C L_C^2 ( e_A - e_C)}{\zeta_A - \zeta_C} - \lambda e_A \right)& A = B\,.
  \end{cases}
\end{equation}
\end{subequations}
Note that $f_t$ and $f_d$ do not appear in the expression for $T_{AB}$. This is expected from equation \eqref{eqn:dGaudin}, where we see that infinitesimal variations of the affine Gaudin parameters corresponding to these two one-forms are given by $(\dd \lambda,\dd \zeta_A,\dd L_A) = - (0,f_t,0) -f_d (\lambda,\zeta_A,L_A) $. Consequentially, $f_t$ encode infinitesimal translations of $\zeta_A$ while $f_d$ encode infinitesimal dilations of $(\lambda,\zeta_A,L_A)$. As argued in the beginning of this subsection, these transformations leave the structure coefficients $F_{ABC}$ invariant, explaining why $f_t$ and $f_d$ decouple from $\dd F_{ABC}$ and thus from $T_{AB}$. Moreover, one sees that the coefficients $f_A$ only appear in differences and thus that $T_{AB}$ is invariant under a shift of all $f_A$'s by the same one-form $c$. Again, this is expected. As argued in the previous paragraph, such a shift (combined with a shift of $f_t$, which does not appear in $T_{AB}$) corresponds to sending $f(z)$ to $f(z)+c\,\varphi(z)^{-1}$ and thus does not change the variation of the twist function. Taking into account these 3 redundancies, we see that the $4N+2$ one-forms contained in $e(z)$ and $f(z)$ induce $4N-1$ independent variations of the structure coefficients, thus matching the number of physical parameters of the model.

It is interesting to note that from the point of view of Lie algebra deformation theory, \eqref{eqn:dF=TF} represents a trivial deformation. It does not change the underlying Lie algebra $\mathfrak{d}$ but merely corresponds to an infinitesimal change of basis in $\mathfrak{d}$. This is natural from the affine Gaudin model construction. Indeed, the variations considered here correspond to infinitesimal transformations along the space $\Pi$ of affine Gaudin parameters. Yet, we know from subsection \ref{sec:poles} that the underlying Lie algebra structure of $\mathfrak{d}$ is in fact independent of the point in $\Pi$. It depends only on the number and order of the poles of the twist function, not on the continuous parameters which enter its expression.

\subsection{Comparison with the RG-flow}
Having understood how an infinitesimal variation of the affine Gaudin parameters acts on the structure coefficients $\Fh_{\Ah\Bh\Ch}$ of the integrable $\EE$-models in the chiral basis, we now want to compare this result with the RG-flow of these theories. The renormalisation of general $\EE$-models was discussed in subsection \ref{sec:RG}, under the assumption that quantum effects preserve the $\EE$-model structure. In particular, we explained there that the whole RG-flow can then be encoded in the variation \eqref{eqn:betaforFh} of the structure coefficients $\Fh_{\Ah\Bh\Ch}$ and recalled the explicit expression of this flow at 1-loop.

In that context, proving the renormalisability of the affine Gaudin models boils down to showing that the RG-flow \eqref{eqn:betaforFh} of $\Fh_{\Ah\Bh\Ch}$ can be fully reabsorbed in a variation of the affine Gaudin parameters\footnote{Assuming that there are no finite quantum corrections to the functional expression of $\Fh_{\Ah\Bh\Ch}$ in terms of these parameters.}, which we recall act on $\Fh_{\Ah\Bh\Ch}$ as \eqref{eqn:dF=TF}. Comparing these two equations, we find that this is the case if and only if there exists a vector field $\beta$ on the parameter space $\Pi$ such that
\begin{equation}\label{eqn:TBeta}
  \iota_\beta \Th_{\Ah}{}^{\Bh} = \betah_{\Ah}{}^{\Bh}\,\bigl|_{\text{int}}\,.
\end{equation}
Here, $\betah_{\Ah}{}^{\Bh}\,\bigl|_{\text{int}}$ denotes the evaluation of the tensor $\betah_{\Ah}{}^{\Bh}$ (appearing in \eqref{eqn:betaforFh}) along the subspace of integrable $\EE$-models (which can then be seen as a function on the space $\Pi$ of affine Gaudin parameters), while $\Th_{\Ah}{}^{\Bh}$ is the one-form on $\Pi$ defined through equation \eqref{eqn:dF=TF} and $\iota_\beta \Th_{\Ah}{}^{\Bh}$ is its interior product with the vector field $\beta$. Before proving that such a $\beta$ exists at one-loop and giving its explicit expression, let us study the above condition in more details.

\paragraph{$\fg$--factorisation.} To analyse \eqref{eqn:TBeta} further, we will exploit some of the general properties of the integrable $\EE$-models under consideration. Indeed, recall that their chiral basis $\lbrace \th_{\Ah} \rbrace$ is labelled by pairs of indices $\Ah = \alpha \,A$, with $\alpha\in\lbrace 1,\dots,\dim\fg\rbrace$ and $A\in\lbrace 1,\dots,2N\rbrace$, and that their structure coefficients $\Fh_{\Ah\Bh\Ch}=f_{\alpha\beta\gamma}\,F_{ABC}$ take a factorised form, where the first factors $f_{\alpha\beta\gamma}$ are the structure constants of the fixed Lie algebra $\fg$ (in an orthonormal basis $\lbrace t_\alpha \rbrace_{\alpha=1}^{\text{dim}\,\fg}$), while the second factors $F_{ABC}$ depend in a specific way on the continuous parameters of the theory. Guided by this observation, let us consider the class of $\EE$-models with a chiral basis labelled by the same pairs of indices $\Ah = \alpha A$ and with structure coefficients also satisfying the factorisation property $\Fh_{\Ah\Bh\Ch}=f_{\alpha\beta\gamma}\,F_{ABC}$, but with general $F_{ABC}$'s.\footnote{Note that we are defining these $\EE$-models directly in the chiral basis, henceforth letting the matrices $\etah_{\Ah\Bh} = \kappa_{\alpha\beta}\,\eta_{AB}$ and $\Eh_{\Ah\Bh}=\kappa_{\alpha\beta}\,\EE_{AB}$ take the same constant diagonalised form as in the integrable case.} The space of such theories, which we refer to as $\fg$--factorised $\EE$-models, is then parametrised by the choice of allowed coefficients $F_{ABC}$ (which should be completely symmetric and such that $\Fh_{\Ah\Bh}{}^{\Ch}=f_{\alpha\beta}{}^\gamma\,F_{AB}{}^C$ satisfy the Jacobi identity). The affine Gaudin models considered in this paper then form a $(4N-1)$--dimensional submanifold in this space, corresponding to the specific choice \eqref{eqn:genfluxes} of coefficients $F_{ABC}$.

Recall that the structure coefficients $\Fh_{\Ah\Bh\Ch}$ in the chiral basis flow according to equation \eqref{eqn:betaforFh}, in terms of the tensor $\betah_{\Ah}{}^{\Bh}$ (if renormalisation preserves the $\EE$-model structure). It is natural to expect that, in an appropriate renormalisation scheme, this tensor is built from covariant contractions of $\etah_{\Ah\Bh}$, $\Eh_{\Ah\Bh}$ and $\Fh_{\Ah\Bh\Ch}$: for $\fg$--factorised models, this can only produce a tensor of the form $\betah_{\Ah}{}^{\Bh} = \delta_\alpha{}^\beta\beta_A{}^B$. Renormalisation then preserves the space of $\fg$--factorised $\EE$-models and induces the following flow on $F_{ABC}$:
\begin{equation}
\frac{\dd F_{ABC}}{\dd \log\mu} = 3 \beta_{(A}{}^{D} F_{BC)D}\,.\vspace{4pt}
\end{equation}

\paragraph{Back to the integrable case.} Let us now come back to the condition \eqref{eqn:TBeta} for renormalisability of the integrable $\EE$-models. Since the latter belong to the class of $\fg$--factorised models, the tensor $\betah_{\Ah}{}^{\Bh}$ appearing on the right-hand side of this equation takes a factorised form $\delta_\alpha{}^\beta\,\beta_A{}^B$, as explained above. Recall moreover from equation \eqref{eqn:defT} that the tensor $\Th_{\Ah}{}^{\Bh}$ appearing on the left-hand side is also factorised as $\delta_\alpha{}^\beta\,T_A{}^B$. The condition for renormalisability then reduces to the simpler form
\begin{equation}\label{eqn:TBeta2}
\iota_\beta T_{AB} = \beta_{AB}\,\bigl|_{\text{int}}\,,
\end{equation}
where we stripped out the $\fg$--factor and lowered the second index for future convenience. As explained in the previous subsection, the tensor $T_{AB}$ is valued in the space of one-forms on $\Pi$ and is given explicitly by equation \eqref{eqn:T}, in terms of the forms $(e_A,f_A)$.

Let us analyse the condition \eqref{eqn:TBeta2}. We first note that $\beta_{AB}=-\beta_{BA}$ is a skew-symmetric tensor. This result comes from subsection \ref{sec:RG}, where we found that any RG-flow that preserves the $\EE$-model structure leads to a skew-symmetric $\betah_{\Ah\Bh}$. Here, the latter takes the factorised form $\kappa_{\alpha\beta}\,\beta_{AB}$, hence the skew-symmetry of $\beta_{AB}$. Yet, the tensor $T_{AB}$ is in general not skew-symmetric. More precisely, its symmetric part is given by equation \eqref{eqn:Ts} in terms of the one-forms $e_1,\dots,e_{2N}$ on $\Pi$. Thus, one finds that equation \eqref{eqn:TBeta2} is possible only if $\iota_\beta e_A = 0$ for $A\in\lbrace 1,\dots,2N\rbrace$. Hence, assuming the hypotheses made in the above discussion, the renormalisation of affine Gaudin models can only lead to an RG-flow of the twist function of the form 
\begin{equation}
  \frac{\dd \;}{\dd \log \mu} \varphi(z) = \partial_z \bigl[ f_\beta(z) \varphi(z) \bigr]\,,
\end{equation}
where $f_\beta(z) = \iota_\beta\,f(z)$.

A comment is in order before we push the analysis further. Recall from subsection \ref{sec:zeroes} that the labels $A$ used in the zeroes basis are naturally decomposed into indices $a\in\lbrace 1,\dots,N\rbrace$, for which $\epsilon_a=+1$, and indices $\ab \in \lbrace 1,\dots, N\rbrace$, for which $\epsilon_{\ab}=-1$, associated with the two chiralities of the model. In the skew-symmetric tensor $\beta_{AB}$ characterising the RG-flow of $\fg$--factorised $\EE$-models, only the components $\beta_{a\bb}=-\beta_{\bb a}$ encode a non-trivial transformation. The remaining components $\beta_{ab}$ and $\beta_{\ab\bb}$ generate double Lorentz transformations and can thus be set to any skew-symmetric expressions without changing the flow of the underlying $\sigma$-model. In practice, this means that we only have to check the condition \eqref{eqn:TBeta2} for $A=a$ and $B=\bb$. Taking into account the expression \eqref{eqn:T} of $T_{AB}$ as well as the fact that we already found $\iota_\beta e_A = 0$ earlier, we then rewrite the renormalisability condition as
\begin{equation}\label{eqn:constraintbeta}
  \iota_{\beta} T_{a\bb} = - \frac{L_a L_{\bb}}{(\zeta_a - \zeta_{\bb})^2} \iota_{\beta} (f_a - f_{\bb}) = \beta_{a\bb}\,\bigl|_{\text{int}}\,.
\end{equation}

\paragraph{Semi-magic squares and curious matrices.} Let us finally discuss the structure underlying the equation \eqref{eqn:constraintbeta}. It is equivalent to
\begin{equation}\label{eqn:AdaptedBeta}
\iota_{\beta} f_a - \iota_{\beta} f_{\bb} = \beta'_{a\bb}\,, \qquad \text{ where } \qquad \beta'_{a\bb} := - \frac{(\zeta_a - \zeta_{\bb} )^2}{L_a L_{\bb}} \beta_{a\bb}\,\bigl|_{\text{int}}
\end{equation}
is called the \textit{adapted $\beta$-tensor}. The matrix appearing on the left-hand side of this condition takes a very specific form. More precisely, it belongs to the linear subspace of $\mathrm{Mat}_{N\times N}$
\begin{equation}
\CM =  \left\{ A = [ u_{a} - v_{\bb} ]\,, \; u,v \in \mathbb{R}^N \right\}\,,
\end{equation}
which is called the space of \textit{curious matrices}~\cite{Mayoral:1996}. Reference~\cite{Mayoral:1996} studies the properties of $\CM$ in detail and provides an interesting alternative characterisation of curious matrices, namely
 \begin{equation}\label{eqn:curiousmatrices}
  \CM =  \left\{ A = [ A_{a\bb} ] \in \mathrm{Mat}_{N\times N} : \sum_{a=1}^N A_{a\overline{\sigma(a)}} \text{ is the same for all $\sigma \in P_N$}\right\}\,,
\end{equation}
where $P_N$ is the set of all permutations of $N$ integers. It also puts forward a relation between curious matrices and \textit{semi-magic squares}. The latter are defined as the $N\times\ N$ matrices whose lines and columns all sum to the same number. They thus form the $(N^2 - 2 N + 2)$-dimensional vector space
\begin{equation}
  \MS = \left\{ A = [ A_{a\bb} ] \in \mathrm{Mat}_{N\times N} : \sum_{a=1}^N A_{a\cb} = \sum_{\bb=1}^N A_{d\bb} \text{ for all $\cb$, $d = 1, \dots , N$}\right\}\,.
\end{equation}
A natural linear subspace of $\MS$, of dimension $(N^2 - 2N + 1)$, corresponds to the case where this number vanishes and is thus given by
\begin{equation}
  \MS(0) = \left\{ A = [ A_{a\bb} ] \in \mathrm{Mat}_{N\times N} : \sum_{a=1}^N A_{a\cb} = \sum_{\bb=1}^N A_{d\bb} = 0 \text{ for all $\cb$, $d = 1, \dots , N$}\right\}\,.
\end{equation}
In this language, the space of curious matrices $\CM$ can be equivalently characterised as the orthogonal subspace of $\MS(0)$ with respect to the standard symmetric bilinear pairing
\begin{equation}
\text{Tr}(AB^T) = \sum_{a,\bb = 1}^N A_{a\bb} B_{a\bb}
\end{equation}
on $\mathrm{Mat}_{N\times N}$, where $B^T$ is the tranpose of $B$. In other words, the space of curious matrices can also be seen as
\begin{equation}
  \CM = \MS(0)^\perp =  \left\{ A \in \mathrm{Mat}_{N\times N} : \text{Tr}(AB^T) = 0 \text{ for all } B \in \MS(0)\,\right\}\,.
\end{equation}

Using the language introduced above, one then finds that the affine Gaudin models are renormalisable if and only if the adapted $\beta$-tensor $\beta'_{a\bb}$, defined in the second equation of \eqref{eqn:AdaptedBeta}, belongs to the space of curious matrices $\CM$. We will show that this is the case at one-loop in the next subsection. In principle, one can hope that this condition, together with the above characterisations of the curious matrices, could help in studying the renormalisability of affine Gaudin models at higher-loop or even at all-loop\footnote{For instance, the characterisation \eqref{eqn:curiousmatrices} of curious matrices shows that the affine Gaudin models are renormalisable if and only if $\sum_{a=1}^N \beta'_{a\overline{\sigma(a)}}$ is the same for all permutations $\sigma\in P_N$. This suggests investigating the properties of symmetries of the RG-flow of integrable $\sigma$-models under permutation of the zeroes $\lbrace\zeta_{\bb}\rbrace$ associated with the same chirality $\epsilon_{\bb}=-1$.}.

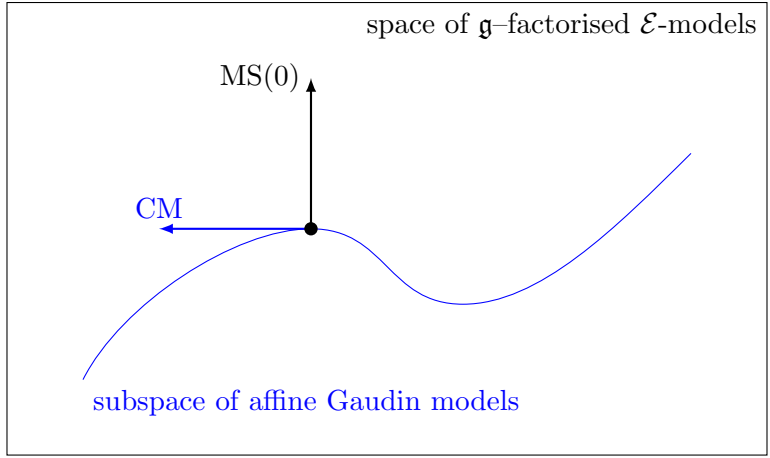
\begin{figure}[t]
\begin{center}
\begin{tikzpicture}
  \draw (0,0) rectangle (10,6);
  \draw[blue] (1,1) node[blue,anchor=north west] {subspace of affine Gaudin models} .. controls (1.5,2) and (3,3) .. (4,3) .. controls (5,3) and (5,2) .. (6,2) .. controls (7,2) and (8,3) .. (9,4);
  \draw[->,thick,blue] (4,3) -- (2,3) node[anchor=south] {$\CM$};
  \draw[->,thick] (4,3) -- (4,5) node[anchor=east] {$\MS(0)$};
  \node[at={(4,3)},minimum size=5pt,fill=black,circle,inner sep=0] {};
  \node[anchor=north east,at={(10,6)}] {space of $\mathfrak{g}$--factorised $\EE$-models};
\end{tikzpicture}
\end{center}
\caption{Embedding of affine Gaudin models in the space of $\mathfrak{g}$--factorised $\EE$-models. For the RG-flow to preserve this subspace, the tensor $\beta'
_{a\bb}$ has to be fully contained in $\CM$, defined in \eqref{eqn:curiousmatrices}, and thus should not have any components in the orthogonal complement $\MS(0)$.}\label{fig:decompcouplings}
\end{figure}

The above formulation also helps the geometric understanding of how affine Gaudin models are embedded into the space of $\fg$--factorised $\EE$-models and how this subspace behaves under a variation of the affine Gaudin parameters. At a given point in this subspace, the RG-flow of the theory is characterised by the adapted $\beta$-tensor $\beta'_{a\bb}$. As depicted in figure~\ref{fig:decompcouplings}, this flow preserves the subspace of affine Gaudin models if this tensor is fully contained in the space of curious matrices $\CM$. On the other hand, if $\beta'_{a\bb}$ has any components in the orthogonal complement $\MS(0)$, the RG-flow will bring the theory to a $\fg$--factorised $\EE$-model which can no longer be interpreted as an affine Gaudin model.

To end this subsection, let us stress again that the various results and ideas discussed above rely on two important assumptions made throughout the reasonning. The first one is that renormalisation preserves the structure of $\EE$-models (see subsection \ref{sec:RG} for details). The second one is that the expression of $F_{ABC}$ in terms of the affine Gaudin parameters, \textit{i.e.} the embedding of affine Gaudin models in the space of $\fg$--factorised $\EE$-models, does not acquire any quantum corrections. Both of these assumptions would require careful further considerations for future investigations at higher-loop.

\subsection{One-loop renormalisability and \texorpdfstring{$\beta$}{beta}-functions}

Expressions for the one- and two-loop $\beta$-functions of $\EE$-models are known \cite{Valent:2009nv,Sfetsos:2009dj,Sfetsos:2009vt,Hassler:2020wnp}. Here, we restrict the discussion to one-loop to keep it simple and demonstrate how the results from the previous sections can be applied. As explained in subsection \ref{sec:RG}, the RG-flow of $\EE$-models in the chiral basis is controlled by the tensor $\betah_{\ah\bhb}$, whose expression at one-loop was given in equation \eqref{eqn:beta-1loop} in terms of certain structure coefficients. We now specialise the discussion to $\fg$-factorised $\EE$-models (see previous subsection), for which these coefficients take the factorised form
\begin{equation}
\Fh_{\ah\chb}{}^{\dh} = f_{\alpha\gamma}{}^\delta\,F_{a\cb}{}^d \qquad \text{ and } \qquad \Fh_{\,\bhb\dh}{}^{\chb} = f_{\beta\delta}{}^\gamma\,F_{\bb d}{}^{\cb}\,.
\end{equation} 
The one-loop tensor \eqref{eqn:beta-1loop} then reads
\begin{equation}
\betah_{\ah\bhb}^{(1)} = - f_{\alpha\gamma}{}^\delta\,f_{\beta\delta}{}^\gamma\,F_{a\cb}{}^d\,F_{\bb d}{}^{\cb}\,.
\end{equation}
We recognise in the first factor the Killing form $f_{\alpha\gamma}{}^\delta\,f_{\beta\delta}{}^\gamma = \text{Tr}\bigl( \text{ad}_{t_\alpha} \text{ad}_{t_\beta} \bigr)$ of $\fg$, which is then proportional to the invariant bilinear form $\kappa_{\alpha\beta}$. More explicitly, let us fix the latter to be $\kappa_{\alpha\beta}=-\text{Tr}(t_\alpha t_\beta)$, where the trace is taken in the fundamental representation of $\fg$ and the minus sign has been introduced so that it is positive-definite for $\fg$ compact. We then have
\begin{equation}
    f_{\alpha\gamma}{}^\delta f_{\beta\delta}{}^\gamma = -2\cg\, \kappa_{\alpha\beta}\,,
\end{equation}
where $\cg$ is the dual Coxeter number of $\fg$. We then find
\begin{equation}
\betah_{\ah\bhb}^{(1)} = 2\cg \, \kappa_{\alpha\beta}\,F_{a\cb}{}^d\,F_{\bb d}{}^{\cb}\,.
\end{equation}
In particular, this is an explicit check at one-loop of a general result argued in the previous subsection, namely that the tensor $\betah_{\ah\bhb}$ of a $\fg$-factorised $\EE$-model takes a factorised form $\kappa_{\alpha\beta}\,\beta_{a\bb}$. We then find the one-loop expression of $\beta_{a\bb}$ to be
\begin{equation}\label{eqn:betaf-1loop}
\beta_{a\bb}^{(1)} =  2\cg \,F_{a\cb}{}^d\,F_{\bb d}{}^{\cb}\,.
\end{equation}

\paragraph{One-loop renormalisation of affine Gaudin models.} We now specialise further to the case of affine Gaudin models, for which the structure constants take the form \eqref{eqn:genfluxes}. Recalling that indices $a,\bb,\cb,d$ are lowered using the bilinear form $\eta_{AB}=\delta_{AB}\,\epsilon_{B}$, with $\epsilon_a=\epsilon_d=+1$ and $\epsilon_{\bb}=\epsilon_{\cb}=-1$ (see subsection \ref{sec:zeroes}), we can perform explicitly the sum over $\cb$ and $d$ in equation \eqref{eqn:betaf-1loop}. This yields
\begin{equation}
\beta_{a\bb}^{(1)}\Bigl|_{\text{int}} = -2\cg\, F_{aa\bb} F_{\bb\bb a} = - 2\cg \frac{L_a L_{\bb}}{(\zeta_a - \zeta_{\bb})^2}\,.
\end{equation}
The adapted $\beta$-tensor defined in \eqref{eqn:AdaptedBeta} then takes the particularly simple form
\begin{equation}
  \beta_{a\bb}^{(1)}{}' = 2\cg\,.
\end{equation}
In the language of the previous subsection, one sees immediately that this is an element of the space of curious matrices $\CM$. In fact, it is further distinguished as it lies in the intersection of curious matrices and semi-magic squares, namely
\begin{equation}
    \beta^{(1)}{}' \in \CM \cap \MS \,.
\end{equation}
This proves that, at least at one-loop, affine Gaudin models are stable under the RG-flow. By comparing the first equation of \eqref{eqn:AdaptedBeta} with the above form of $\beta_{a\bb}^{(1)}{}'$, one sees that the vector field $\beta$ is given at one-loop by
\begin{equation}
\iota_{\beta^{(1)}} f_a = + \cg =  \cg\,\epsilon_a \qquad \text{ and } \qquad \iota_{\beta^{(1)}}  f_{\bb} = - \cg =  \cg\,\epsilon_{\bb}\,,
\end{equation}
while we recall that $\iota_{\beta^{(1)}}e_A=0$ (see previous subsection). Plugging this vector field in equation \eqref{eqn:dphi}, we eventually also extract the one-loop RG-flow of the twist function:
\begin{equation}
  \frac{\dd \varphi(z)}{\dd\log\mu} = \alpha' \,\partial_z \left[ f^{(1)}(z) \varphi(z) \right]\,,
\end{equation}
with
\begin{equation}
f^{(1)}(z) = \iota_{\beta^{(1)}} f(z) = \cg \sum_{A=1}^{2N} \frac{L^2_A}{z-\zeta_A} + f_t^{(1)} + f_d^{(1)}\,z \,.
\end{equation}
We recall that $f_t^{(1)} = \iota_{\beta^{(1)}} f_t$ and $ f_d^{(1)} = \iota_{\beta^{(1)}} f_d$ are left unfixed as they encode dilations and rescalings of the spectral parameter and thus leave the affine Gaudin model invariant (see subsection \ref{sec:var}). This shows that the conjecture made in \cite{Delduc:2020vxy} and proven in \cite{Hassler:2020xyj} for the case of single poles is true for twist functions with arbitrary pole structure and simple zeros in the complex plane (and a double pole at infinity).

\section{Conclusion and perspectives}
The goal of this paper was to study the renormalisation of a large class of integrable $\sigma$-models obtained from the formalism of affine Gaudin models~\cite{Levin:2001nm,Feigin:2007mr,Vicedo:2017cge} or equivalently from 4-dimensional Chern-Simons theory~\cite{Costello:2019tri,Vicedo:2019dej}. In particular, we established the 1-loop renormalisability of these theories and computed explicitly the RG-flow of their defining parameters, using the formalism of $\EE$-models. Moreover, we showed that this RG-flow can be recast in a very simple and compact way in terms of the so-called twist function of the model, thus confirming a conjecture first proposed in~\cite{Delduc:2020vxy} and proved in the particular case of twist functions with simple poles in~\cite{Hassler:2020xyj}. Finally, we exhibited the structure underlying the renormalisability of these theories as being related to curious matrices and semi-magic squares.

A crucial role in our analysis is played by the zeroes of the twist function, which are parts of the natural parameters describing affine Gaudin models. For the theories considered in the present article, these zeroes are assumed to be simple (corresponding to having simple poles in the Lax connection). It would be interesting to extend the construction of these models and the analysis of their RG-flow to the case of twist functions with higher order zeroes.

Another natural direction for generalisation is to study the renormalisation of gauged affine Gaudin models~\cite{Vicedo:2017cge,Lacroix:2019xeh,benini:2020skc,Arutyunov:2020sdo,Liniado:2023uoo}, which for instance include integrable $\sigma$-models on symmetric spaces and their deformations. We expect that the methods and results of this paper naturally extend to this more general family of models (in particular, we hope that this question can be approached using the work \cite{Severa:2018pag} on the renormalisation of degenerate $\EE$-models/dressing cosets).

Various other results and conjectures on the 1-loop renormalisation of integrable $\sigma$-models were discussed in recent works. This includes a formulation of the RG-flow using integrals of the twist function in~\cite{Derryberry:2021rne} and a universal formula for the divergences of these models in terms of their Lax connection in~\cite{Levine:2022hpv}\footnote{See also the recent work~\cite{Levine:2023wvt} for a proof that this universal formula matches the divergences computed from the 4-dimensional Chern-Simons theory with disorder defects.}. It would be interesting to investigate the relation between these different approaches and the one developed in the present paper.

Finally, a quite interesting but challenging perspective of this work is its extension beyond the first order in the quantum loop expansion. The study of higher-loop renormalisation of integrable $\sigma$-models has been an active domain of research in the past few years: see for instance~\cite{Hoare:2019ark,Hoare:2019mcc,Georgiou:2019nbz,Levine:2021fof}. In particular, these works show that renormalisability beyond one-loop generally requires the addition of quantum corrections to the geometry of the model. It would be interesting to understand these corrections in the language of $\EE$-models used in the present article and to study the higher-loop RG-flow of the twist function.

\section*{Acknowledgements}
We would like to thank Anders Heide Wallberg and Nat Levine for interesting discussions. The work of FH is supported by the SONATA BIS grant 2021/42/E/ST2/00304 from the National Science Centre (NCN), Polen. The work of SL is supported by Dr. Max R\"ossler, the Walter Haefner Foundation and the ETH Z\"urich Foundation. BV gratefully acknowledge the support of the Leverhulme Trust through a Leverhulme Research Project Grant (RPG-2021-154).

\appendix

\section{Generalised frame field and isotropic basis}
\label{app:Frame}
As explained in subsection \ref{sec:sigma} of the main text, the generalised frames $\sEh_{\Ah}{}^I$ provide the link between an $\EE$-model and the corresponding \s-models. Thus, it is worth reviewing how they are constructed. Most important in the construction is that we deal with two Lie algebras, $\dlie$ and $\hlie$, where the latter is a maximally isotropic subalgebra of the former. To emphasise this point, one splits the generators of $\dlie$ into 
\begin{equation}
  \th_{\Ah} = \big( \th_{\ah}\,, \th^{\,\ah} \big)
\end{equation}
and requires that the second half, $\th^{\,\ah}$ spans $\hlie$. An immediate consequence is $\langle \th^{\,\ah}, \th^{\,\bh} \rangle = 0$. Up to a redefinition of the complement basis $\{ \th_{\ah} \}$, one can always bring $\etah_{\Ah\Bh}$ into the canonical form
\begin{equation}\label{eqn:etahiso}
  \etah_{\Ah\Bh} = \begin{pmatrix}
    0 & \delta_{\ah}{}^{\bh} \\
    \delta^{\ah}{}_{\bh} & 0
  \end{pmatrix}\,.
\end{equation}
Similar to the $\etah$-metric, the structure coefficients decompose into the four independent contributions
\begin{equation}
  \Fh_{\ah\bh\ch}\,, \quad
  \Fh_{\ah\bh}{}^{\ch}\,, \quad
  \Fh_{\ah}{}^{\bh\ch}\,, \quad \text{and} \quad
  \Fh^{\,\ah\bh\ch} = 0\,,
\end{equation}
where the last one vanishes because $\hlie$ is assumed to be a subalgebra of $\dlie$. We use the fixed form of $\etah_{\Ah\Bh}$ in \eqref{eqn:etahiso} and $\Fh^{\,\ah\bh\ch}=0$ as the defining properties of what we call isotropic basis. These conditions are not sufficient to uniquely identify an isotropic basis and we will discuss the freedom in its choice later in this appendix.

For the moment, let us consider a fixed isotropic basis $\big( \th_{\ah}\,, \th^{\,\ah} \big)$ and discuss how the generalised frame field is constructed from this data. Recall that the space we are interested in is the quotient $M=\HH \backslash \DD$. In what follows, we assume that we have made a smooth choice of coset representative of every $m\in \HH \backslash \DD$, so that we can see $m$ as an element of $\DD$. We will need the right-invariant one-form
\begin{equation}
  \dd m m^{-1} = \th_{\ah} v^{\ah}{}_i \dd x^i + \th^{\,\ah} A_{\ah i} \dd x^i 
\end{equation}
and the adjoint action
\begin{equation}
  \Mh_{\Ah}{}^{\Bh} \th_{\Bh} = m t_{\Ah} m^{-1}\,.
\end{equation}
In these terms, the generalised frame field is then defined as
\begin{equation}\label{eqn:paramgenframe}
  \sEh_{\Ah}{}^I = \Mh_{\Ah}{}^{\Bh} \begin{pmatrix}
    \vh_{\bh}{}^i & \vh_{\bh}{}^j B_{ji} \\
    0 & v^{\bh}{}_i
  \end{pmatrix}\,.
\end{equation}
Here, $\vh_{\ah}{}^i$ denote the components of the dual vector fields corresponding to the one-forms $v^{\ah} = v^{\ah}{}_i \dd x^i$ with the defining property $\iota_{\vh_{\ah}} v^{\bh} = \vh_{\ah}{}^i v^{\bh}{}_i = \delta_{\ah}{}^{\bh}$. Moreover, we need the two form $B$-field
\begin{equation}
  B = \frac12 B_{ij}\, \dd x^i \wedge \dd x^j = \frac12 v^{\ah} \wedge A_{\ah} + B_{\mathrm{WZW}}
\end{equation}
with
\begin{equation}
  \dd B_{\mathrm{WZW}} = - \frac12 \langle \dd m m^{-1}, [ \dd m m^{-1}, \dd m m^{-1} ] \rangle\,.
\end{equation}
Note that $B_{\mathrm{WZW}}$ is usually not globally well-defined, but can always be obtained patch-wise and then glued together by $B$-field gauge transformations.

For completeness, let us now come back to the question of the non-uniqueness of the isotropic basis $\big( \th_{\ah}\,, \th^{\,\ah} \big)$. To explore this, we note that $\etah_{\Ah\Bh}$ is the invariant metric of the Lie group O($\Dh$,$\Dh$) with $\Dh = \frac12 \dim \DD = \dim \HH$. Hence, any new basis $\big( \th'_{\ah}\,, \th'^{\,\ah} \big)$ obtained from $\big( \th_{\ah}\,, \th^{\,\ah} \big)$ by an O($\Dh$,$\Dh$) transformation will leave $\etah_{\Ah\Bh}$ invariant. However, for $\big( \th'_{\ah}\,, \th'^{\,\ah} \big)$ to also define an isotropic basis, one still has to check the second constraint on the structure constants. To this end, we parameterise O($\Dh$,$\Dh$) transformations which are connected to the identity as
\begin{equation}
  \begin{aligned}
    \th'_{\ah} &= \Ah_{\ah}{}^{\bh}\, \th_{\bh} + \Bh_{\ah\bh}\, \th^{\bh} \\
    \th'^{\ah} &= \Ah^{\,\ah}{}_{\bh}\, \th^{\bh} + \betah^{\,\ah\bh}\, \th_{\bh}
  \end{aligned}
\end{equation}
with $\Ah_{\ah}{}^{\ch} \Ah^{\,\bh}{}_{\ch} = \delta_{\ah}{}^{\bh}$, $\Bh_{\ah\bh} = -\Bh_{\bh\ah}$ and $\betah_{\ah\bh} = -\betah_{\bh\ah}$. By construction,  $\etah_{\Ah\Bh}$ is invariant under this transformation. We note that the coefficients $\Ah_{\ah}{}^{\bh}$ essentially amount to a change of the basis $\{ \th^{\,\bh} \}$ of $\hlie$. Similarly, the coefficients $\Bh_{\ah\bh}$ encode a change in the choice of basis $\{ \th_{\bh} \}$ of the complement of $\hlie$ in $\dlie$, without modifying $\hlie$ itself. For any transformation with $\betah^{\,\ah\bh}=0$, $\th'_{\ah}$ then still span the same maximally isotropic subalgebra $\hlie$. The most interesting transformations are then the ones encoded in $\betah^{\,\ah\bh}$, which correspond to a change in the choice of $\hlie$. By construction, the new subspace $\hlie'$ spanned by $\{ \th'^{\,\bh} \}$ is always isotropic; moreover, we find that it is a closed subalgebra if and only if
\begin{equation}\label{eqn:goodbetatransform}
  \Fh'^{\,\ah\bh\ch} = \betah^{\,\ah\dh} \betah^{\,\bh\eh} \betah^{\,\ch\fh} \Fh_{\dh\eh\fh} + 3 \betah^{\,[\ah|\dh} \betah^{\,|\bh|\eh} \Ah^{\,|\ch\,]}{}_{\fh} \Fh_{\dh\eh}{}^{\fh} + 3 \betah^{\,[\ah|\dh} \Ah^{\,|\bh}{}_{\eh} \Ah^{\,\ch\,]}{}_{\fh} \Fh_{\dh}{}^{\eh\fh} = 0\,.
\end{equation}
The constraint \eqref{eqn:goodbetatransform} can in general have non-trivial solutions, leading to new choices of maximally isotropic subalgebras and thus to different \s-model realisations in the language of $\EE$-models. This idea has been introduced in \cite{Borsato:2021vfy} to study the space of T-dual models.

\section{Changing from the poles basis to the zeroes basis}
\label{app:ChangeBasis}
In this appendix, we perform the explicit transformation between the poles basis and the zeroes basis of the double algebra underlying the integrable $\mathcal{E}$-models of Section \ref{sec:integrability}. Recall that in both of these bases, the structure constants and the non-degenerate pairing take a factorised form. More precisely, the first factor, common to both bases, is labelled by indices of the Lie algebra $\fg$ while the second one is labelled by either poles or zeroes. To perform the change of basis, it is thus enough to focus on this second factor. In the case of the poles basis, we use indices of the form $\widetilde A = [i,p]$, where $i\in\lbrace 1,\dots,M\rbrace$ labels the position $z_i$ of the poles and $p\in\lbrace 0,\dots,n_i-1\rbrace$ labels their order. In the zeroes basis, we use indices $A\in\lbrace 1,\cdots,2N \rbrace$ labelling the (simple) zeroes $\zeta_A$, with $2N=\sum_{i=1}^M n_i$. The relation between the two bases is explicitly given by the (rescaled) confluent Cauchy matrix with entries \cite{Lacroix:2020flf}
\begin{equation}\label{eqn:Cauchy}
\mathcal{C}^A\!{}_{[i,p]}= -\frac{\epsilon_A\,L_A}{(\zeta_A-z_i)^{p+1}}\,,
\end{equation}
where we recall that $L_A=\bigl|\varphi'(\zeta_A)\bigr|^{-1/2}$. Indeed, the $\EE$-model currents $J^{\alpha [i,p]}_\sigma(\sigma)$ associated with the poles -- see equation \eqref{eqn:GaudinLax} -- are related to the ones $\J^{\alpha A}_\sigma(\sigma)$ associated with the zeroes -- see equation \eqref{eqn:ourLax} -- by
\begin{equation}
\J^{\alpha A}_\sigma(\sigma) = \sum_{i=1}^M \sum_{p=0}^{n_i-1} \mathcal{C}^A\!{}_{[i,p]}\, J^{\alpha [i,p]}_\sigma(\sigma)\,.
\end{equation}

\paragraph{Invariant pairing.} Let us now consider the invariant pairing. Its behaviour under the change of basis can be read from~\cite[Lemmas 4.2 and 4.3]{Lacroix:2020flf}, Here, we rederive this result by elementary computations. In the poles basis, the pairing is given by the tensor $\widetilde \eta^{\widetilde A\widetilde B}$ defined in equation \eqref{eqn:tildeEta}. Applying the change of basis \eqref{eqn:Cauchy}, we find
\begin{align}
\mathcal{C}^A\!{}_{\widetilde A} \, \mathcal{C}^B\!{}_{\widetilde B}\; \widetilde \eta^{\widetilde A\widetilde B} &= \epsilon_A\epsilon_B\,L_A L_B\sum_{i=1}^M \sum_{p,q=0}^{n_i-1} \frac{\ell_{[i,p+q]}}{(\zeta_A-z_i)^{p+1}(\zeta_B-z_i)^{q+1}}\, \label{eqn:AppEta}\\
&=\epsilon_A\epsilon_B\,L_A L_B \sum_{i=1}^M \sum_{r=0}^{n_i-1} \ell_{[i,r]} \sum_{p=0}^r \frac{1}{(\zeta_A-z_i)^{p+1}(\zeta_B-z_i)^{r-p+1}}\,, \notag
\end{align}
where in the last equality we have changed the sum over $(p,q)$ to a sum over $(r,p)$ with $r=p+q$, using the fact that $\ell_{[i,r]}=0$ if $r\geq n_i$. We now need to distinguish whether $A$ coincides with $B$ or not. We start with the case $A\neq B$, using the identity
\begin{equation*}
\sum_{p=0}^r \frac{1}{(\zeta_A-z_i)^{p+1}(\zeta_B-z_i)^{r-p+1}} = \frac{1}{\zeta_B-\zeta_A} \left( \frac{1}{(\zeta_A-z_i)^{r+1}} - \frac{1}{(\zeta_B-z_i)^{r+1}} \right)\,.
\end{equation*}
Substituting this into the above equation and using the expression \eqref{eqn:GaudinLax} of $\varphi(z)$, one easily finds
\begin{equation}
\mathcal{C}^A\!{}_{\widetilde A} \, \mathcal{C}^B\!{}_{\widetilde B}\; \widetilde \eta^{\widetilde A\widetilde B} = - \epsilon_A\epsilon_B\,L_A L_B\, \frac{\varphi(\zeta_A)-\varphi(\zeta_B)}{\zeta_A-\zeta_B} = 0\,,
\end{equation}
since by definition the $\zeta_A$'s are the zeroes of $\varphi(z)$. Let us now consider equation \eqref{eqn:AppEta} in the case $A=B$. Recalling that $\epsilon_A^2=1$, we find
\begin{equation}
\mathcal{C}^A\!{}_{\widetilde A} \, \mathcal{C}^A\!{}_{\widetilde B}\; \widetilde \eta^{\widetilde A\widetilde B} = L_A^2 \sum_{i=1}^M \sum_{r=0}^{n_i-1} \frac{(r+1)\ell_{[i,r]}}{(\zeta_A-z_i)^{r+2}} = -L_A^2 \,\varphi'(\zeta_A) = \epsilon_A\,,
\end{equation}
where in the last equality we used $\varphi'(\zeta_A)=-\epsilon_A\,L_A^{-2}$. In summary, we thus proved $\mathcal{C}^A\!{}_{\widetilde A} \, \mathcal{C}^B\!{}_{\widetilde B}\; \widetilde \eta^{\widetilde A\widetilde B}=\epsilon_A\,\delta_{AB}$, which then agrees with the tensor $\eta^{AB}$ defined in equation \eqref{eqn:eta}.

\paragraph{Structure constants.} The main non-trivial computation we have to perform is the change of basis of the structure constants. In the poles basis, these constants $\widetilde F^{\widetilde A\widetilde B}{}_{\widetilde C}$ are given by equation \eqref{tildeF}. To simplify the computation, we will use the structure constants with all indices raised, by contracting the third index with $\widetilde\eta$. Using the expression \eqref{eqn:tildeEta} of the latter, we get
\begin{equation}
\widetilde F^{\,[i,p]\,[j,q]\,[k,r]} = \delta_{ij}\delta_{ik} \ell_{[i,p+q+r]}\,.
\end{equation}
Performing the change of basis, we then obtain
\begin{align}
&\mathcal{C}^A\!{}_{\widetilde A}\, \mathcal{C}^B\!{}_{\widetilde B}\,\mathcal{C}^C\!{}_{\widetilde C}\;\widetilde F^{\widetilde A\widetilde B\widetilde C} \\ 
&\hspace{30pt} = -\epsilon_A\epsilon_B\epsilon_C\,L_A L_B L_C \sum_{i=1}^M \sum_{p,q,r=0}^{n_i-1} \frac{\ell_{[i,p+q+r]}}{(\zeta_A-z_i)^{p+1}(\zeta_B-z_i)^{q+1}(\zeta_C-z_i)^{r+1}}\,. \notag
\end{align}
To compute this expression, it will be useful to introduce the function of three complex variables
{\small\begin{equation}
\mathcal{F}(w_1,w_2,w_3) = -\sum_{i=1}^M \sum_{s=0}^{n_i-1} \ell_{[i,s]}\!\! \sum_{\substack{p,q,r=0 \\ p+q+r=s}}^{n_i-1} \frac{1}{(w_1-z_i)^{p+1}(w_2-z_i)^{q+1}(w_3-z_i)^{r+1}}\,,
\end{equation}}so that
\begin{equation}
\mathcal{C}^A\!{}_{\widetilde A}\, \mathcal{C}^B\!{}_{\widetilde B}\,\mathcal{C}^C\!{}_{\widetilde C}\;\widetilde F^{\widetilde A\widetilde B\widetilde C}  = \epsilon_A\epsilon_B\epsilon_C\,L_A L_B L_C\, \mathcal{F}(\zeta_A,\zeta_B,\zeta_C)\,.
\end{equation}
To simplify this function, we apply the identity
\begin{align}
&\!\!\sum_{\substack{p,q,r=0 \\ p+q+r=s}}^{n_i-1} \!\frac{1}{(w_1\!-\!z_i)^{p+1}(w_2\!-\!z_i)^{q+1}(w_3\!-\!z_i)^{r+1}} = \frac{1}{(w_1\!-\!w_2)(w_1\!-\!w_3)} \frac{1}{(w_1\!-\!z_i)^{s+1}} \\
&\hspace{25pt} + \frac{1}{(w_2\!-\!w_1)(w_2\!-\!w_3)} \frac{1}{(w_2\!-\!z_i)^{s+1}} + \frac{1}{(w_3\!-\!w_1)(w_3\!-\!w_2)} \frac{1}{(w_3\!-\!z_i)^{s+1}}\,. \notag
\end{align}
Using also the expression \eqref{eqn:GaudinLax} of $\varphi(z)$, we then find
\begin{align}
\mathcal{F}(w_1,w_2,w_3) &= -\left( \frac{\varphi(w_1)}{(w_1-w_2)(w_1-w_3)} + \frac{\varphi(w_2)}{(w_2-w_1)(w_2-w_3)}  + \frac{\varphi(w_3)}{(w_3-w_1)(w_3-w_2)} \right). \notag
\end{align}
We now have to evaluate this function at $(w_1,w_2,w_3)=(\zeta_A,\zeta_B,\zeta_C)$. Due to the presence of terms $w_i-w_j$ in the denominator, we have to be careful when some of the indices $A,B,C$ coincide. The simplest situation is thus when the three indices are distinct, in which case we clearly find $\mathcal{F}(\zeta_A,\zeta_B,\zeta_C)=0$ since the $\zeta_A$'s are the zeroes of $\varphi(z)$, hence $\mathcal{C}^A\!{}_{\widetilde A}\, \mathcal{C}^B\!{}_{\widetilde B}\,\mathcal{C}^C\!{}_{\widetilde C}\;\widetilde F^{\widetilde A\widetilde B\widetilde C}=0$. In the case of two coinciding labels, a more careful analysis shows
\begin{equation}
\mathcal{F}(\zeta_A,\zeta_A,\zeta_B) = -\frac{\varphi'(\zeta_A)}{\zeta_A-\zeta_B} = \frac{\epsilon_A L_A^{-2}}{\zeta_A-\zeta_B}\,.
\end{equation}
We then get
\begin{equation}
\mathcal{C}^A\!{}_{\widetilde A}\, \mathcal{C}^A\!{}_{\widetilde B}\,\mathcal{C}^B\!{}_{\widetilde C}\;\widetilde F^{\widetilde A\widetilde B\widetilde C}  = \epsilon_B\,L_A^2 L_B \mathcal{F}(\zeta_A,\zeta_A,\zeta_B) = \frac{\epsilon_A\epsilon_B L_B}{\zeta_A-\zeta_B}\,.
\end{equation}
Finally, the analysis of the case with all equal indices gives
\begin{equation}
\mathcal{F}(\zeta_A,\zeta_A,\zeta_A) = - \frac{1}{2} \varphi''(\zeta_A)\,.
\end{equation}
A direct computation using equation \eqref{eqn:phiinv} then shows that
\begin{equation*}
\mathcal{C}^A\!{}_{\widetilde A}\, \mathcal{C}^A\!{}_{\widetilde B}\,\mathcal{C}^A\!{}_{\widetilde C}\;\widetilde F^{\widetilde A\widetilde B\widetilde C}  = \epsilon_A\,L_A^3\, \mathcal{F}(\zeta_A,\zeta_A,\zeta_A) = \frac{\epsilon_A}{L_A} \left( \lambda - \sum_{B \neq A} \frac{\epsilon_B\,L_B^2}{\zeta_A-\zeta_B}\right)\,.
\end{equation*}
We thus conclude that, in all these cases, the tensor $\mathcal{C}^A\!{}_{\widetilde A}\, \mathcal{C}^B\!{}_{\widetilde B}\,\mathcal{C}^C\!{}_{\widetilde C}\;\widetilde F^{\widetilde A\widetilde B\widetilde C} $ coincides with $F^{ABC}$ defined through equation \eqref{eqn:genfluxes} in the main text.\footnote{Note that the equation \eqref{eqn:genfluxes} gives the coefficients $F_{ABC}$ with all indices lowered. The expression found above for $\mathcal{C}^A\!{}_{\widetilde A}\, \mathcal{C}^B\!{}_{\widetilde B}\,\mathcal{C}^C\!{}_{\widetilde C}\;\widetilde F^{\widetilde A\widetilde B\widetilde C}$ has to be compared with $F^{ABC}=\epsilon_A\,\epsilon_B\,\epsilon_C\,F_{ABC}$, where we raised the indices using $\eta^{AB}=\epsilon_A\,\delta^{AB}$. We then find complete agreement with the results of this appendix.}

\bibliography{literature}
   
\bibliographystyle{JHEP}

\end{document}